\documentclass[aps,prx,superscriptaddress,twocolumn,showpacs]{revtex4-1}
\usepackage[table]{xcolor}
\usepackage{pgf}
\usepackage[utf8]{inputenc}
\usepackage{amsmath}
\usepackage{braket}
\usepackage{amssymb}
\usepackage{amsmath}
\usepackage{bbold}
\usepackage{enumitem} 
\usepackage{appendix}
\usepackage{graphicx}
\usepackage{overpic}
\usepackage{dsfont}
\usepackage{tikz}
\usepackage[pdftex,bookmarks,colorlinks]{hyperref}
\usepackage{hyperref}
\usepackage{booktabs}

\usepackage{ulem}

\newcommand{\Tr}[1]{\operatorname{Tr} #1}
\newcommand{\mean}[1]{\left\langle #1\right\rangle}
\newcommand{\nep}{\textrm{e}}
\newcommand{\nb}{{\boldsymbol{n}}}
\newcommand{\ha}{{\hat{a}}}

\newcommand{\hn}{{\hat{n}}}

\newcommand{\bp}{{\boldsymbol{\psi}}}
\newcommand{\be}{{\boldsymbol{\eta}}}

\begin{document}
\title{Many-body dynamical localization in the kicked Bose-Hubbard chain}
\author{Michele Fava}
\affiliation{Rudolf Peierls Centre for Theoretical Physics, Clarendon Laboratory, University of Oxford, Oxford OX1 3PU, UK}

\author{Rosario Fazio}
\affiliation{Abdus Salam ICTP, Strada Costiera 11, I-34151 Trieste, Italy}
\affiliation{Dipartimento di Fisica, Universit\`a di Napoli "Federico II", Monte S. Angelo, I-80126 Napoli, Italy}

\author{Angelo Russomanno}
\affiliation{Abdus Salam ICTP, Strada Costiera 11, I-34151 Trieste, Italy}
\affiliation{Max-Planck-Institut f\"ur Physik Komplexer Systeme, N\"othnitzer Strasse 38, D-01187, Dresden, Germany}

\begin{abstract}
{We provide evidence that a clean kicked Bose-Hubbard model exhibits a many-body dynamically localized phase. This phase {shows} ergodicity breaking {up to the largest sizes we were able to consider}. {We argue that this property persists in the limit of large size}. {The Floquet states violate eigenstate thermalization and then the asymptotic value of local observables depends on the initial state and is not thermal. This implies that the system does not generically heat up to infinite temperature, for almost all the initial states.} Differently from many-body localization here the entanglement entropy linearly increases in time. This increase corresponds to space-delocalized Floquet states which are nevertheless localized across specific subsectors of the Hilbert space: In this way the system is prevented from randomly exploring all the Hilbert space and does not thermalize.}
\end{abstract}

\maketitle
\section{Introduction}
%
Although the relation between classical chaos and thermalization in many-body systems is a very well established subject, its quantum counterpart is still mysterious in many respects. There is a strong evidence that in thermalizing quantum systems the eigenstates are locally thermal (Eigenstate Thermalization Hypothesis -- ETH~\cite{Deutsch_PRA91,Sred_PRE94,Rigol_Nat,ikeda}) but it is not possible to predict {\it a priori} if this thermalization occurs. {The physical process behind thermalization is different from the classical case (no more chaotic and ergodic trajectories in phase space but locally thermal eigenstates), indeed classically thermalizing systems can behave in a regular and non-ergodic way in the quantum regime. An important example} is many-body localization (MBL) where disorder makes a quantum chaotic and thermalizing system to become integrable and space-localized~\cite{Bloch_2019}. 

In this work we focus on periodically-driven systems {which have attracted a lot of attention in last years (see Ref.~\cite{Eckardt_RMP_17} for a review).}
{In particular, many efforts have been devoted to understand the effect of quantum mechanics on energy absorption from the external driving~\cite{Russomanno_PRL12,Russomanno_EPL15,Buco_arXiv15,Emanuele_2014:preprint,mori_prl16,mori_AnnP,Knappa_preprint15,nature_knap}}. In this context, an ergodic ``thermalizing'' behaviour coincides with an indefinite heating up: Because of the absence of energy conservation, thermalization can occur only at $T=\infty$. From a classical perspective, efficient energy absorption occurs
when the driving is resonant with some natural frequencies of the system and chaos
develops around resonances~\cite{Berry_regirr78:proceeding,Lichtenberg,chiri_vov}. Trajectories diffuse in the chaotic part of the phase space and this gives rise to energy absorption.  Moving to the quantum case, thermalization and energy absorption occur when the eigenstates of the stroboscopic dynamics (the Floquet states) obey ETH: They are random delocalized states, locally equivalent to the $T=\infty$ thermal ensemble~\cite{Lazarides_PRE14,Dalessio_PRX14}, and the level-spacing distribution of the corresponding eigenvalues (the Floquet quasienergies) is of the circular-orthogonal-ensemble form~\cite{Dalessio_PRX14}. 
 In this context, an important question is understanding when the quantum-classical correspondence disappears and quantum interference breaks chaos and energy absorption.

{The simplest example of this sort is the phenomenon of dynamical localization~\cite{Boris:rotor,Casati:rotor}. It occurs in the dynamics of a periodically-kicked single-degree-of-freedom system, the so-called kicked rotor. Classically this non-integrable model can show a fully-chaotic behaviour where it absorbs energy from the driving diffusively in time and without a bound. Going to the quantum regime, interference deeply affects the energy dynamics: After a transient the energy absorption is arrested (this is the reason for the name dynamical localization). A question arisen in the last years is if it is possible to see dynamical localization in the thermodynamic limit, in {\it homogeneous} systems where there is no MBL. Research in this sense has been performed on the many-body version of the quantum kicked rotor~\cite{Notarnicola_PRB18}. There,  persistence of dynamical localization was observed for a finite number of rotors, but this phenomenon always seemed to disappear in the thermodynamic limit.
There are two examples of existence of dynamical localization in the thermodynamic limit ({\it many-body dynamical localization} -- MBDL) for non-integrable models, {\it i)} in a system which at high energies tends to an integrable model~\cite{rozenbaum_e} and {\it ii)} in a kicked Lieb-Liniger gas, for initial states with small  energy~\cite{rylands}.} 

{Refs.~\cite{rozenbaum_e,rylands} are important steps towards the understanding of many-body dynamical localization, however there are many important aspects of this phenomenon that need to be clarified. It is not clear if the many-body dynamical
localization occurs for any initial state. It should be understood that, if many-body dynamical localization is a phase
affecting all the spectrum, where the Floquet states violate
ETH, all of them are locally nonthermal and then there is no thermalization, despite the initial conditions. It would be
of great importance to characterize this form of ergodicity
breaking from other already known forms, as MBL. 
}

Here we address these questions by discussing the dynamics of a clean periodically-kicked Bose-Hubbard model which shows clear signatures of dynamical localization in the large-size limit. The system is non linear and apparently non integrable because there is not any known extensive set of commuting local integrals of motion. Nevertheless there is a phase where the dynamics is not chaotic and the system does not heat up to $T=\infty$, {and this property appears to persist} in the large-size limit. {In this regime the overlap with the initial state keeps values of order one and shows finite-size revivals at times linearly scaling with the system size. This behaviour is in sharp contrast with the ergodic case where there is a very fast decrease to values exponentially small in the system size and no revival.} We remark that these dynamical behaviours are observed for different initial conditions, also with a finite initial-energy density: We check that what we find is a phase affecting all the spectrum where all the Floquet states break eigenstate thermalization. We find also that increasing the kicking strength there is a transition to ETH; {this transition involves the properties of all the Floquet states, so it is a true eigenstate phase transition~\cite{eigenstate}}. 

This many-body dynamically localized (MBDL) phase is different from MBL. We show this by looking at the properties of entanglement, in particular of the half-chain entanglement entropy. This quantity increases logarithmically in time in the MBL regime, while we see that it increases linearly in time in the dynamically-localized case, {as it occurs in clean quantum integrable systems~\cite{Alba_PNAS,Russomanno_2016,Alba_SciPostPhys_17,Cala_linea}}. 
 Entanglement properties are very different also between dynamically-localized and ergodic phases. The difference is in the distribution of the Floquet-states entanglement entropies and in the way their average and fluctuations scale with the system size. 
In the end, we compare our quantum results with those obtained from a classical truncated-Wigner analysis, showing that many-body dynamical localization disappears as soon as quantum coherence is broken, {substituted by a chaotic and diffusive behaviour}. As appropriate for dynamical localization, this is a purely quantum phenomenon with no classical analog. {We wish to remark that our quantum results are based on numerics and we can do our simulations for $L\leq 19$ at the best. Although our calculations hint at a  many-body dynamical localization, we emphasize that extrapolations to the thermodynamic limit should be taken with care.}

The paper is organized as follows. In Section~\ref{modmet:sec} we introduce the model we study, the observables we consider and the methods we use. In Section~\ref{evidences:sec} we discuss all the numerical evidences and the system-size scalings leading us to conclude that there is dynamical localization and persists in the thermodynamic limit. In particular, in Sec.~\ref{entanglement:sec} we discuss the behaviour of the entanglement entropy and we show that it behaves very differently in the MBDL, in the MBL and in the ETH phase. It is crucial to distinguish MBDL from MBL, being the behaviour of local observables similar in the two cases, and the time-behaviour of entanglement entropy allows to do that. In Sec.~\ref{fock:sec} we discuss the delocalization properties of the Floquet states in the Hilbert space in the dynamically-localized phase. In Sec.~\ref{sec:TWA} we compare our results with a truncated-Wigner (TWA) analysis showing the purely quantum nature of dynamical localization which disappears in the classical limit. In Section~\ref{conc:sec} we make our conclusions. In the Appendixes we discuss important technical details.

\section{Model and methods} \label{modmet:sec}
\subsection{The model} \label{model:sec}
We consider a periodically-kicked Bose-Hubbard model 
\begin{align} \label{Hamour:eqn}
  \hat{H}(t)&=\hat{V}+\delta_\tau(t)\hat{K}\quad{\rm with}\nonumber\\
  \hat{V}&\equiv\frac{U}{2}\sum_{j=1}^L\hat{n}_j(\hat{n}_j-1)\,,\nonumber\\
  \hat{K}&\equiv J\sum_{j=1}^L\left(\hat{a}_j^\dagger\hat{a}_{j+1}+{\rm H.~c.}\right)\,,
\end{align}
where $L$ is the system size, $\hat{a}_j$ are bosonic operators and $\hat{n}_j\equiv \hat{a}_j^\dagger\hat{a}_{j}$ are number operators {and -- using the same notation as in~\cite{chiri_vov} -- $\delta_\tau(t)=\sum_m\delta(t-m\tau)$ is a periodic delta kicking with period $\tau$. $\hat{V}$ is the undriven interaction part of the Hamiltonian, while $\hat{K}$ is the hopping part coupled to the kicking. In the rest of the paper we will take $\hbar=1$ and we will fix the period $\tau=1$ and the interaction strength $U=1$. In order to explore the different dynamical behaviours we will modify the hopping strength $J$. It is important to mention that the version of this model with static hopping (where to $\delta_\tau(t)$ is replaced 1) is known to undergo a Kosterlitz-Thouless quantum phase transition in its ground state, from Mott insulator to superfluid~\cite{fisher,elstner}. Recently the Kibble-Zurek scaling of excitations has been analyzed for a ramping through this quantum-phase-transition point~\cite{zurek}. The quench dynamics in a Bose-Hubbard model with static hopping has been numerically analyzed in~\cite{kollath,kollath1} and there are clear signs of a transition between a thermalizing and a regular behaviour (although, as in our case, the system sizes are limited and no extrapolation to the thermodynamic limit is rigorously possible). The many-body localization dynamics with static hopping and on-site disorder has been experimentally analyzed in~\cite{julian,Lukin256}.}

It is easy to see that the total number of bosons operator $\hat{N}\equiv\sum_j\hat{n}_j$ is conserved. Defining the filling factor $\nu\equiv \mean{\hat{N}}/L$, it is possible to see that the Hilbert space dimension is $\dim\mathcal{H}=\binom{L(\nu+1)-1}{L-1}$ {(if $L\nu$ is an integer)}. So we have a finite Hilbert space. This implies that, in order to see if there is thermalization, we have to compare the behaviour of the observables with a {\it finite} $T=\infty$ value. This simplifies the analysis compared for instance to Ref.~\cite{Notarnicola_PRB18} where the $T=\infty$ condition coincided with infinite energy and one had to study the energy increase in time with different truncations of the Hilbert space. Here we do not need any truncation. In our analysis we will restrict to the case $\nu=1$ (one boson per site, {on average}) and we will give to the basis of the simultaneous eigenstates of all the $\hat{n}_j$ operators ($\forall\,j$) the name ``Fock basis'' (we will denote it as $\left\lbrace\ket{\nb}\right\rbrace$). It is immediate to see that this is also the basis of the eigenstates of the undriven part of the Hamiltonian $\hat{V}$.

Throughout the work we will consider periodic boundary conditions (unless otherwise specified), so the system has the translation and the inversion symmetries. Therefore, choosing initial states invariant under these symmetry operations, the dynamics restricts to the Hilbert subspace $\mathcal{H}_S$ fully symmetric under these symmetry operations. This fact restricts  the dimension of the interesting Hilbert subspace, so we can perform full exact diagonalizations for system sizes up to $L= 11$ (we can also go beyond using appropriate truncation schemes as we explain later). In particular, the fully-symmetric subspace has a value of the dimension smaller than the full Hilbert space dimension of a factor $\sim 2L$ (see Appendix~\ref{symm:app} for more details). If we do not specify otherwise, numerical results in the figures are obtained through full exact diagonalization.

\subsection{Numerical methods} \label{nummeth:sec}
{For larger system sizes we can study the dynamics by resorting to other numerical methods and approximations. The most versatile of such methods is the Krylov method (implemented in {\sc Expokit}~\cite{EXPOKIT}) which allows to directly compute the time evolution of a state. Similarly, we have also used the tDMRG~\cite{Schollwock_rev} algorithm (implemented through the ITensor Library~\cite{ITensor}). This method is particularly suited to study the dynamics from a translationally-invariant product state when there is a regime of many-body dynamical localization and the entanglement entropy increases linearly but with a small rate (we will better discuss this point in Sec.~\ref{entanglement:sec}). Finally, {for exact diagonalization and Krylov technique} deep in the localized regime, we can exploit the fact that a Fock state will evolve in time to states that are a linear combination of only relatively few other Fock states. In particular, given two Fock states $\ket{\nb}$ and $\ket{\nb'}$, we define the distance $\Delta(\nb,\nb')=\min_\sigma\left[\sum_j \left|n_j-n'_{\sigma(j)}\right|/2\right]$, with $\sigma$ denoting a generic permutation of the $L$ sites. So to study the dynamics starting from a Fock state $\ket{\nb_0}$, we either compute the time evolution or infinite-time averages (see Sec.~\ref{dyn:sec}) working in the Hilbert space spanned by the set of $\ket{\nb}$ s.t. $\Delta(\nb_0,\nb) \leq \Delta_T$. For the purpose of understanding how reliable this approximation is we define the projector $\Pi_\mathcal{B}$ on the border of the truncation $\mathcal{B}=\mbox{Span}\left(\ket{\nb}\, s.t.\, \Delta(\nb_0,\nb) = \Delta_T \right)$ and we use its expectation value as an estimate of the error due to the Hilbert space truncation.}
\subsection{Dynamics, quantities and observables} \label{dyn:sec}
In order to numerically analyze if this system is thermalizing or not, we consider the evolution of the undriven (interacting) part of the Hamiltonian $\hat{V}$ (see  Eq.~\eqref{Hamour:eqn}). We define $\ket{\psi(0)}$ as the initial state of the dynamics and we study the unitary dynamics with Hamiltonian Eq.~\eqref{Hamour:eqn}. We focus on discrete stroboscopic times $m\tau^-$ (just before each kick). Here the state is given by
\begin{equation} \label{statot:eqn}
  \ket{\psi(m\tau^-)}=\left[\hat{U}(\tau^-,0^-)\right]^m\ket{\psi(0)}
\end{equation}
where $\hat{U}(\tau^-,0^-)$ is the time-evolution operator over one period expressed as
\begin{equation} \label{Uop:eqn}
  \hat{U}(\tau^-,0^-)=\nep^{-i\hat{V}\tau}\nep^{-i\hat{K}}\,
\end{equation}
in terms of the components of the Hamiltonian Eq.~\eqref{Hamour:eqn}. We define the stroboscopic expectation of the operator $\hat{V}$ as
\begin{equation}
 \label{overt:eqn}
 V(t)\equiv\bra{\psi(t)}\hat{V}\ket{\psi(t)}\quad{\rm with}\quad t =m\tau^-
\end{equation}
and we define also the infinite-time stroboscopic average of this quantity as
\begin{equation} \label{overv:eqn}
  \overline{V}\equiv\lim_{\mathcal{T}\to\infty}\frac{1}{\mathcal{T}}\sum_{m=0}^{\mathcal{T}}V(m\tau^-)\,.
\end{equation}
With these ingredients, we can define a quantity showing if the system thermalizes at infinite time or not. For that purpose, we start from a suggestion of Ref.~\cite{PONTE_2015,Dalessio_PRX14}, and consider the following quantity
\begin{equation} \label{epsilon:eqn}
  \epsilon\equiv\frac{\overline{V}-V(0)}{V_{T=\infty}-V(0)}\,.
\end{equation}
Here we have defined $V(0)\equiv\bra{\psi(0)}\hat{V}\ket{\psi(0)}$ as the initial expectation of $\hat{V}$, $V_{T=\infty}=\frac{1}{\dim\mathcal{H}_S}\Tr_S\left(\hat{V}\right)$ as the thermal expectation at $T=\infty$.~\cite{Nota1}
%
%
From one side $\epsilon=1$ corresponds to $T=\infty$ thermalization ($\overline{V}=V_{T=\infty}$); from the other there is MBDL if $\epsilon$ tends to a value smaller than one as the system size is increased ($\overline{V}<V_{T=\infty}$ going towards large values of $L$).
 In the paper we will use also the stroboscopic instantaneous version of Eq.~\eqref{epsilon:eqn} defined as
\begin{equation} \label{epsilont:eqn}
  \epsilon(t)\equiv\frac{{V}(t)-V(0)}{V_{T=\infty}-V(0)}\quad\text{with}\quad t=m\tau^-\,.
\end{equation}

When the size is too large and we have to resort to Krylov or tDMRG (see Sec.~\ref{nummeth:sec}), we must approximate the infinite-time average of Eq.~\eqref{overv:eqn} with a long (but finite) time average. On the opposite, when $L\leq 10$ we can use exact diagonalization and apply Floquet theory to compute the exact value of this quantity. Floquet theory~\cite{Sambe_PRA73,Shirley_PR65} states that there exists a basis of solutions of the Schr\"odinger equation which are periodic up to a phase factor. They are obtained as the eigenvectors of the time-evolution operator over one period Eq.~\eqref{Uop:eqn}, $\hat{U}(\tau^-,0^-)\ket{\phi_\alpha}=\nep^{-i\tau\mu_\alpha}\ket{\phi_\alpha}$. The eigenvectors $\ket{\phi_\alpha}$ are the Floquet states and the factors $\mu_\alpha$ in the phases are the corresponding quasienergies. We can expand the time-evolved state Eq.~\eqref{statot:eqn} in the Floquet basis as $\ket{\psi(m\tau^-)}=\sum_\alpha R_\alpha \nep^{-im\tau\mu_\alpha}\ket{\phi_\alpha}$ ($R_\alpha\equiv\bra{\phi_\alpha}\left.\psi(0)\right\rangle$) and substitute this expansion in Eq.~\eqref{overv:eqn}. Assuming that there are no degeneracies in the Floquet spectrum (as it is the case when both $U$ and $J$ are nonvanishing and we have restricted to the fully-symmetric subspace), performing the infinite time-average kills all the oscillating off-diagonal terms~\cite{Dalessio_PRX14} and we get the Floquet-diagonal-ensemble average~\cite{Russomanno_PRL12}
\begin{equation} \label{valpha:eqn}
  \overline{V} = \sum_\alpha |R_\alpha|^2V_\alpha\quad{\rm with}\quad V_\alpha\equiv\bra{\phi_\alpha}\hat{V}\ket{\phi_\alpha}\,.
\end{equation}
%
The Floquet states and the expectations $V_\alpha$ will play an important role in our subsequent analysis. It will be relevant to study their localization properties in the basis of the eigenstates of the undriven part of the Hamiltonian, in order to do that we will consider the Inverse participation ratio~\cite{Edwards_JPC72} of a Floquet state in this basis, defined as
\begin{equation} \label{IPR:eqn}
  {\rm IPR}_\alpha=\sum_{\nb}|\Braket{\nb|\phi_\alpha}|^4\,.
\end{equation}
Another important quantity we will study is the half-chain entanglement entropy~\cite{Nielsen}, in order to explore the entanglement properties of the Floquet states and of the time-evolved state. Given a state $\ket{\psi}$ and a bipartition of the system in two subsystems, $A$ and $B$, each $L/2$-sites long, the half-chain entanglement entropy is defined as
\begin{equation} \label{entropy:eqn}
  S_{L/2}=-\Tr_A[\hat{\rho}_A\log\hat{\rho}_A]\quad{\rm with}\quad \hat{\rho}_A=\Tr_B[\ket{\psi}\bra{\psi}]\,.
\end{equation}
%

{We will have to perform averages over the Floquet spectrum; we define
\begin{equation}
  \mean{(\cdots)}\equiv\frac{1}{\dim\mathcal{H}_s}\sum_{\alpha=1}^{\dim\mathcal{H}_s}(\cdots)_\alpha
\end{equation}
as the average over the Floquet states and
\begin{equation}
  \operatorname{std}(\cdots)=\left[\mean{(\cdots)}^2-\mean{(\cdots)^2}\right]^{1/2}
\end{equation}
as the corresponding standard deviation.}
\section{Evidences for many-body dynamical localization} \label{evidences:sec}
\subsection{Energy dynamics}
\begin{figure}
 \begin{center}
  \begin{tabular}{c}
   \includegraphics[width=7cm]{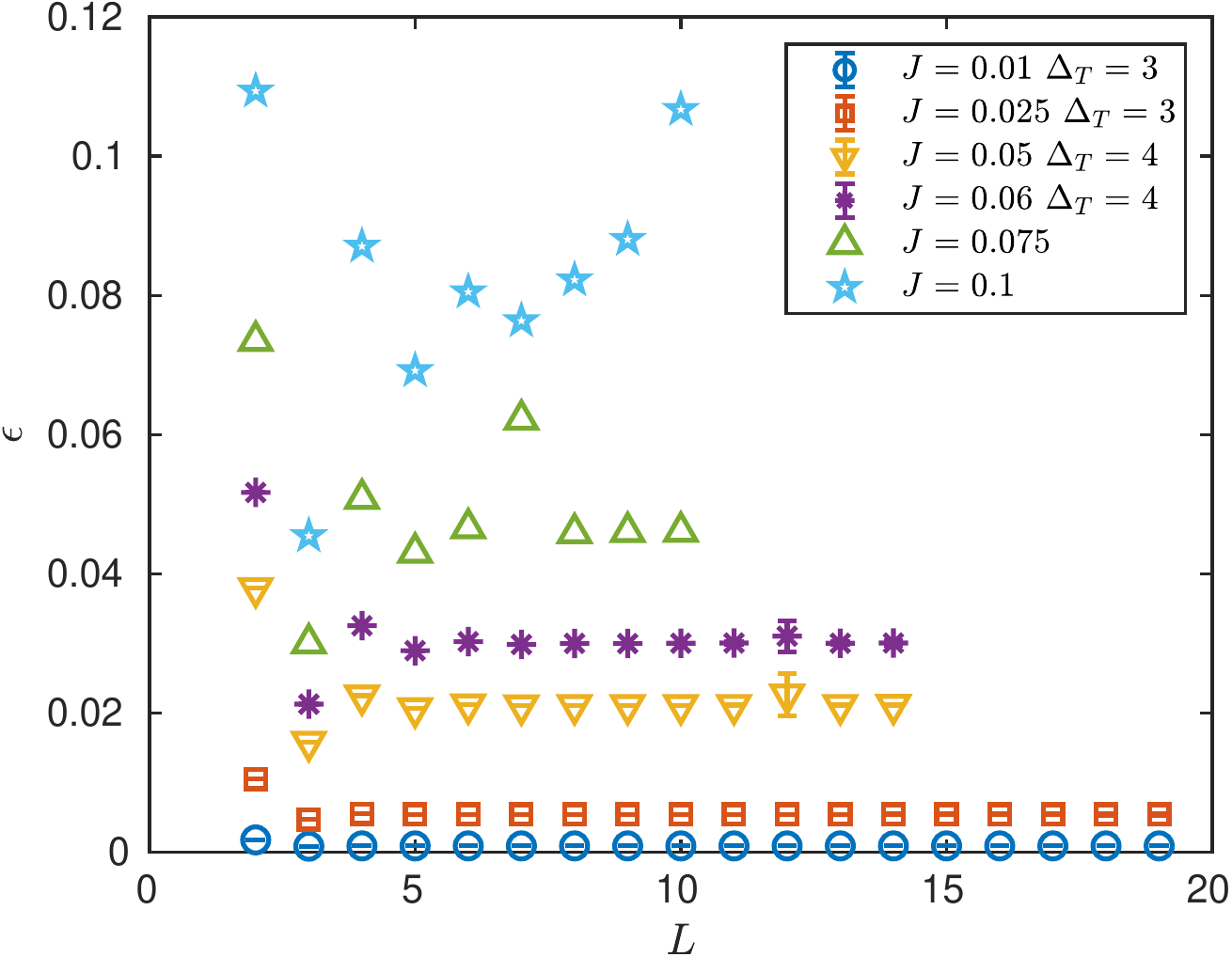}\\
   \includegraphics[width=7cm]{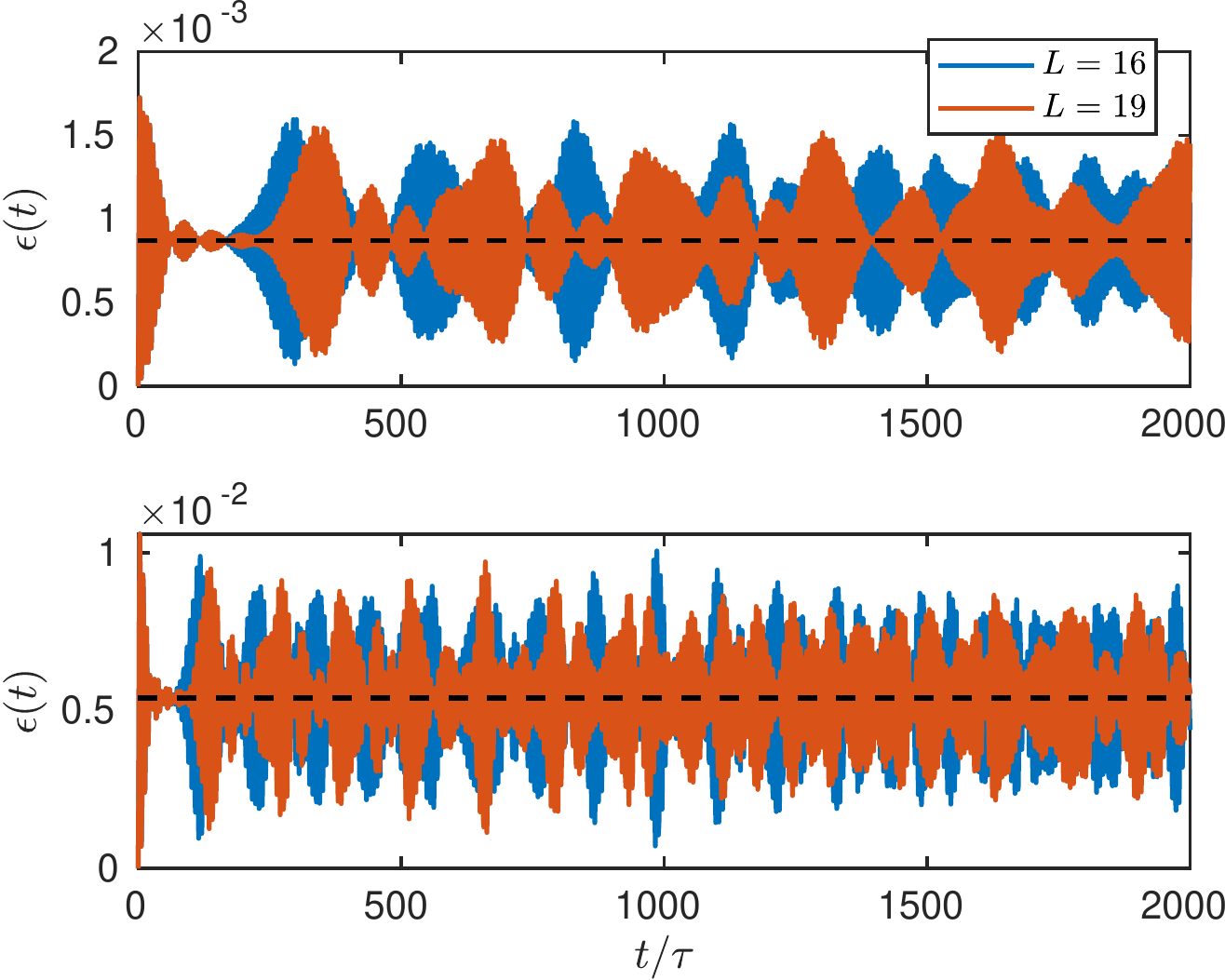}
  \end{tabular}
 \end{center}
 \caption{{(Upper panel) $\epsilon$ versus $L$. For $J\leq 0.06$, notice the convergence to a value much smaller than the thermal value at $T=\infty$, $\epsilon_{T=\infty}=1$.} Data up to $L=10$ are obtained through exact diagonalization in the full Hilbert space.
 The truncation distance $\Delta_T$ employed beyond $L=10$ is indicated in the legend.
 {For points computed using an Hilbert space truncation, the error bar denotes the average value of the expectation of the projector $\Pi_{\mathcal{B}}$.}
 For $\Delta_T=3$ and $L>15$ and $\Delta_T=4$ and $L>12$, we have employed Krylov method. When we use this method, we average $\bar V$ over the first $2000$ kicks.
 (Central and lower panel) $\epsilon(t)$ versus $t$ in two MBDL cases treated with the Krylov method, as elucidate above. We take $J=0.01$ in the central panel and $J=0.025$ in the lower one.
 }
 \label{eps_vs_L:fig}
\end{figure}

In this section we are going to show to the reader all the numerical evidences supporting the existence of many-body dynamical localization. The first quantity we focus on is $\epsilon$ defined in Eq.~\eqref{epsilon:eqn}. First of all, let us take as initial state the uniform tensor product $\ket{\psi(0)}=\ket{n_1=1}\otimes\ket{n_2=1}\otimes\ldots\otimes\ket{n_L=1}$ and study the dependence of $\epsilon$ on $L$ in the upper panel of Fig.~\ref{eps_vs_L:fig}. When $J$ is small enough ($J\lesssim 0.05$), $\epsilon$ reaches a value much smaller than 1, therefore very far from the thermal $T=\infty$ value. Moreover, for $L\leq 15$ this is an infinite-time average (with a truncation for $L>10$), so it is really a dynamical localization and not a prethermal behaviour. Although for $L>10$ the averages are over a finite (but long) time, it would be very strange that a prethermal behaviour started setting from a certain size on, at least we are not aware that this phenomenon has ever been observed in the literature. We are corroborated in this conviction by the behaviour of $\epsilon(t)$ versus $t$ in the MBDL case (see central and lower panels of Fig.~\ref{eps_vs_L:fig}). We see oscillations around an average value that decrease in amplitude until a finite-size revival sets in, at a time which increases with the system size (we discuss more about finite-size revivals in Sec.~\ref{overlap:sec}). After that, oscillations of $\epsilon(t)$ go on around a non-thermal average value much smaller than one (in Fig.~\ref{eps_vs_L:fig} we mark this average by a black dashed line). This behaviour of $\epsilon(t)$ is exactly the same observed for the stroboscopic dynamics of local observables in driven integrable systems~\cite{Russomanno_PRL12} where thermalization is never attained. In Appendix~\ref{02:app} we consider a different initial state and we see the same behaviour of $\epsilon$ versus $L$; this fact strongly suggests that what we see is not a low-energy phenomenon but a property involving all the spectrum.
%
%

%
%

\begin{figure}
 \begin{center}
  \begin{tabular}{c}
   \includegraphics[width=7cm]{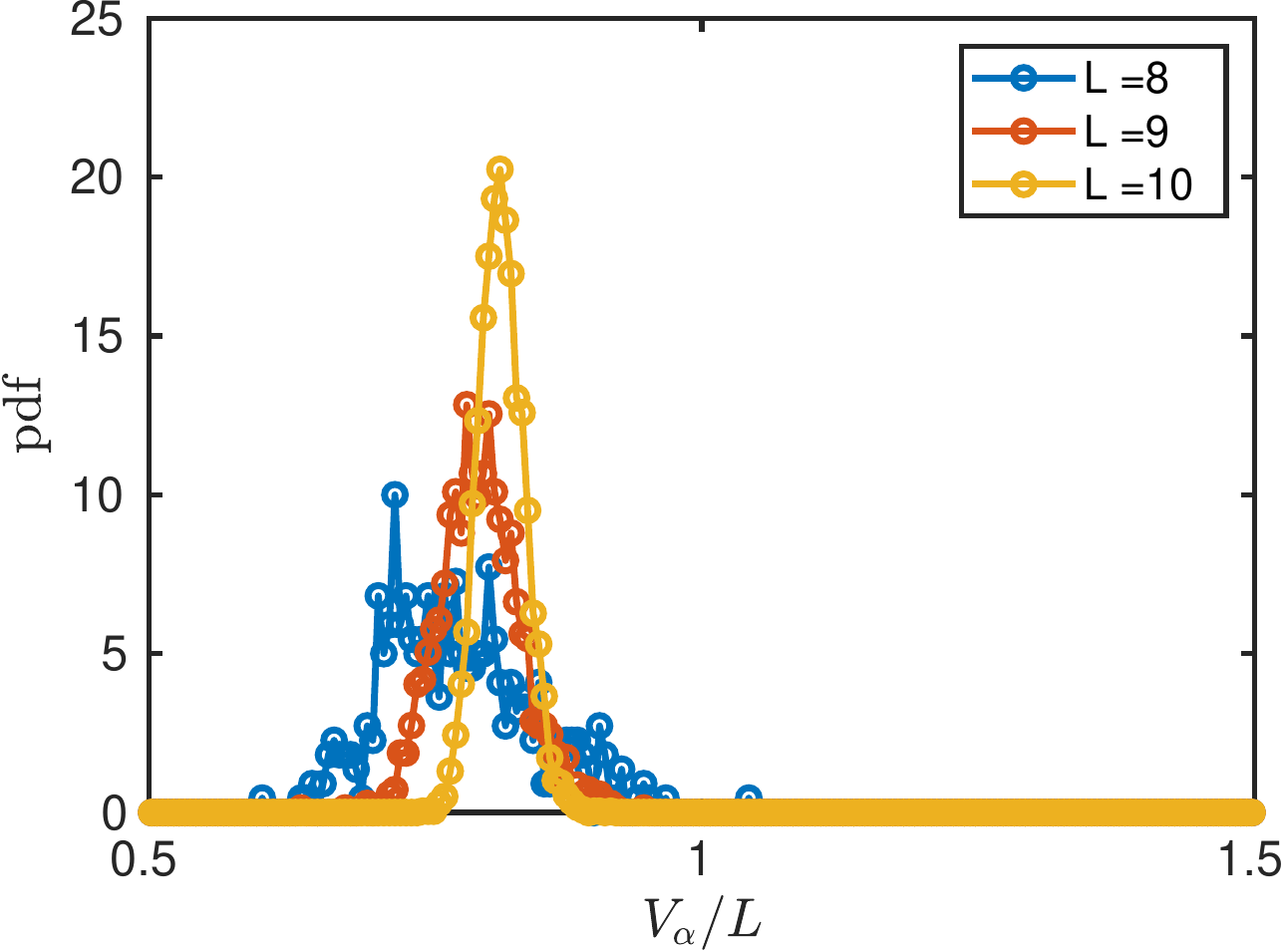}\\
   \includegraphics[width=7cm]{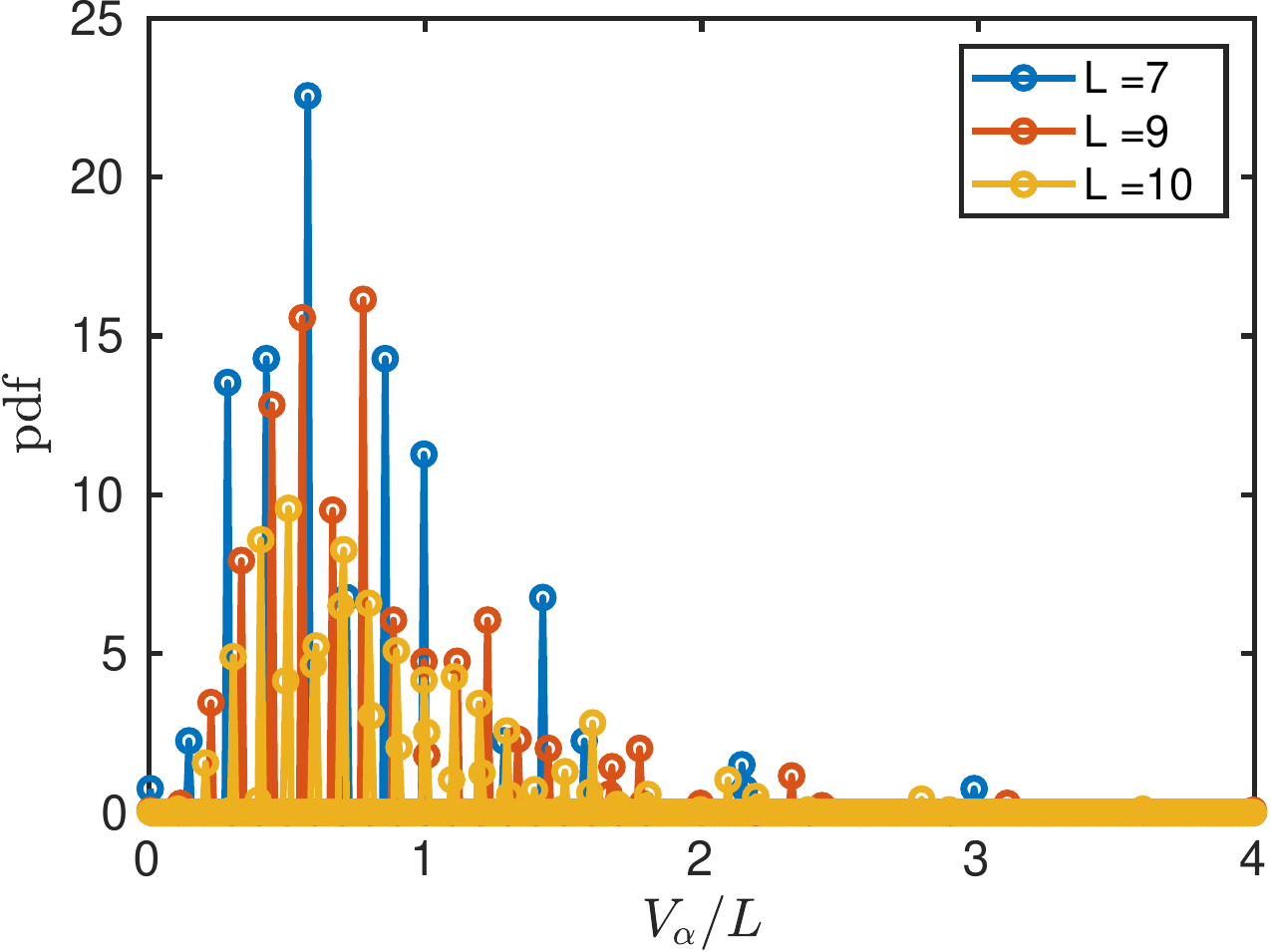}
  \end{tabular}
 \end{center}
 \caption{Upper panel, distribution of $V_\alpha/L$ for a thermalizing case ($J=0.5$). Lower panel, distribution of $V_\alpha/L$ for a dynamically localized case ($J=0.01$).}
 \label{pdf:fig}
\end{figure}

This picture is corroborated by the analysis of the distributions of $V_\alpha$ defined in Eq.~\eqref{valpha:eqn}. First of all, let us show to the reader the behaviour of the distributions of $V_\alpha$ in a thermalizing case and in a case showing dynamical localization (see Fig.~\ref{pdf:fig}). In the former case (upper panel) we have what we expect from eigenstate thermalization: a smooth distribution which gets narrower and narrower around the $T=\infty$ value as the system size is increased. In the latter case, we find something very different. We have {broad} distributions which stay {broad} as the system size is increased (later we will clarify what do the spikes mean).

\begin{figure}
 \begin{center}
  \begin{tabular}{c}
   \includegraphics[width=7cm]{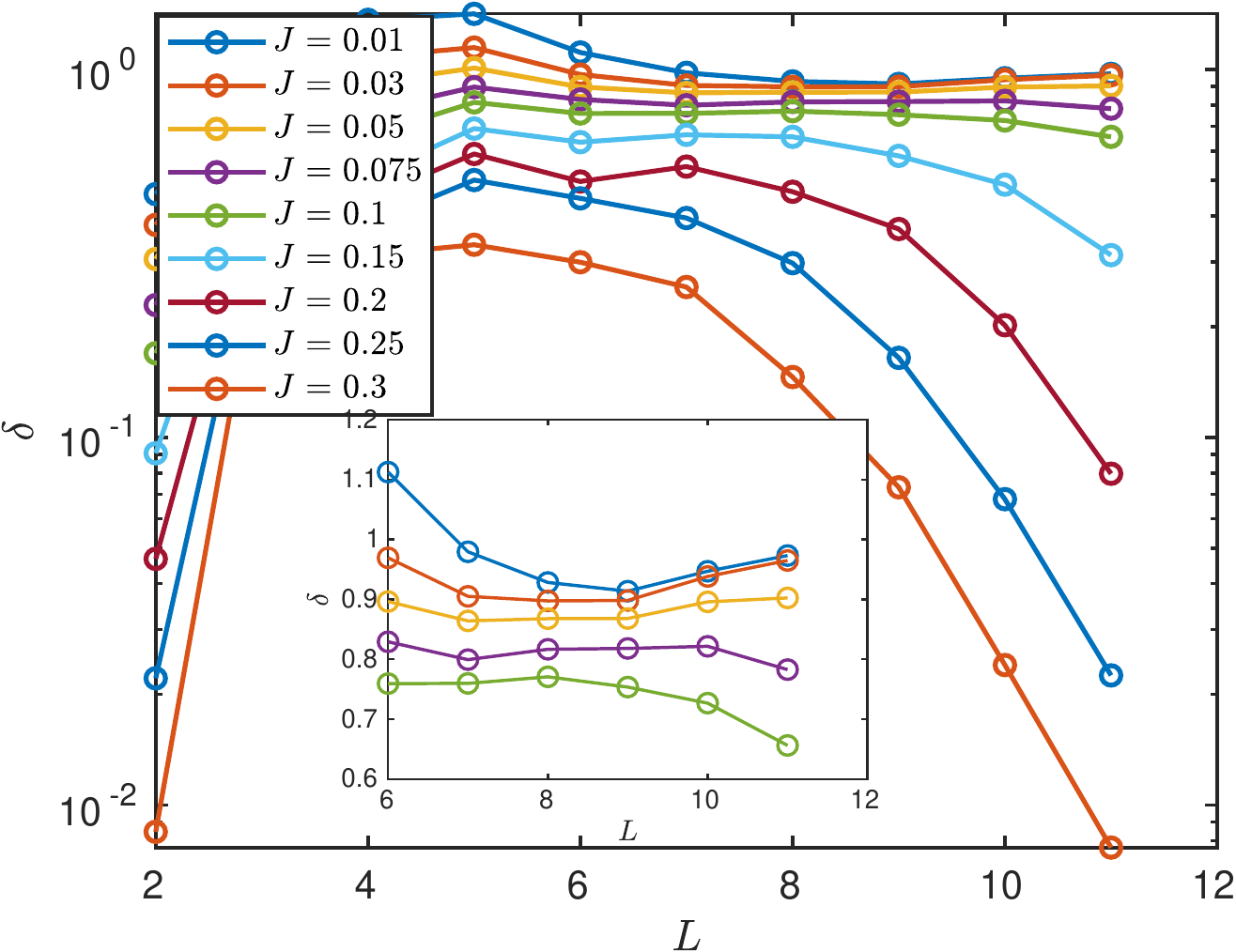}\\
  \end{tabular}
 \end{center}
 \caption{Dependence on $L$ of $\delta\equiv |\mean{V}-\exp\mean{\log V}|$. Notice the bending upwards for {$J\lesssim 0.05$ and the bending downwards when $J\gtrsim 0.075$ which develops in a clear decay for $J\gtrsim0.15$. The first behaviour is consistent with MBDL, the second with eigenstate thermalization. The inset shows a magnification of the plot for the values of $J$ around the change of scaling behaviour.}}
 \label{exp_log_scaling:fig}
\end{figure}

A more quantitative analysis is possible considering the scaling with $L$ of the quantity
%
\begin{equation}
  \delta\equiv|\mean{V}-\exp\mean{\log V}|\,.
\end{equation}
We expect to see a very different scaling in case of thermalization and in case of dynamical localization. In the former case there should be ETH and all the Floquet states should behave locally as the $T=\infty$ density matrix (typicality), up to fluctuations vanishing in the thermodynamic limit. Therefore, the $V_\alpha$-distribution should become more and more narrow as the system size increases: In the thermodynamic limit $L\to\infty$, the mean of $V_\alpha$ must equal the exponential of the mean of $\log V_\alpha$ and then $\delta$ should scale to 0. On the opposite, in the case of dynamical localization, the distributions stay very {broad}: if we see that $\delta$ does not scale to 0 with the increasing system size but tends to a finite value, we can be sure that there is dynamical localization in the large-$L$ limit. We study this scaling in Fig.~\ref{exp_log_scaling:fig} and the results are remarkable. For $J\gtrsim 0.15$ we can see that {for $L$ above a certain threshold, $\delta$ starts decreasing down to 0 with $L$ in a clear way: If this behaviour persists for larger $L$, it leads to ETH and thermalization in the thermodynamic limit. The bending downwards is already visible for $J=0.075$.} For $J\lesssim 0.05$ there is no scaling to 0, on the opposite $\delta$ attains a finite value and bends upwards (lower panel of Fig.~\ref{exp_log_scaling:fig}). Up to $L=11$ the system is dynamically localized and this is a property of all the Floquet spectrum, not only of some specific initial states. For sure we cannot exclude from these data that the curves for $J\lesssim 0.05$ start bending downwards from $L=12$ onward but unfortunately our numerics cannot bring us further. {These results are consistent with those in the upper panel of Fig.~\ref{eps_vs_L:fig}, where we see convergence in $L$ of $\epsilon$ to a value smaller than one for $J\leq 0.06$. We remark that the system sizes we can access to are not large enough to clearly find where is the transition point (and if the transition persists for larger sizes). We can only find a crossover regime between $J=0.06$ and $J=0.1$. Nevertheless, our aim is studying the behaviour of the system in the MBDL and thermalizing phase and not pinpoint the precise value of the transition. So, the main message of Fig.~\ref{exp_log_scaling:fig} is that the behaviour of $\delta$ versus $L$ is qualitatively and quantitatively different in the two phases. }



\subsection{Overlap} \label{overlap:sec}
\begin{figure}
 \begin{center}
   \begin{tabular}{c}
   	\includegraphics[width=7cm]{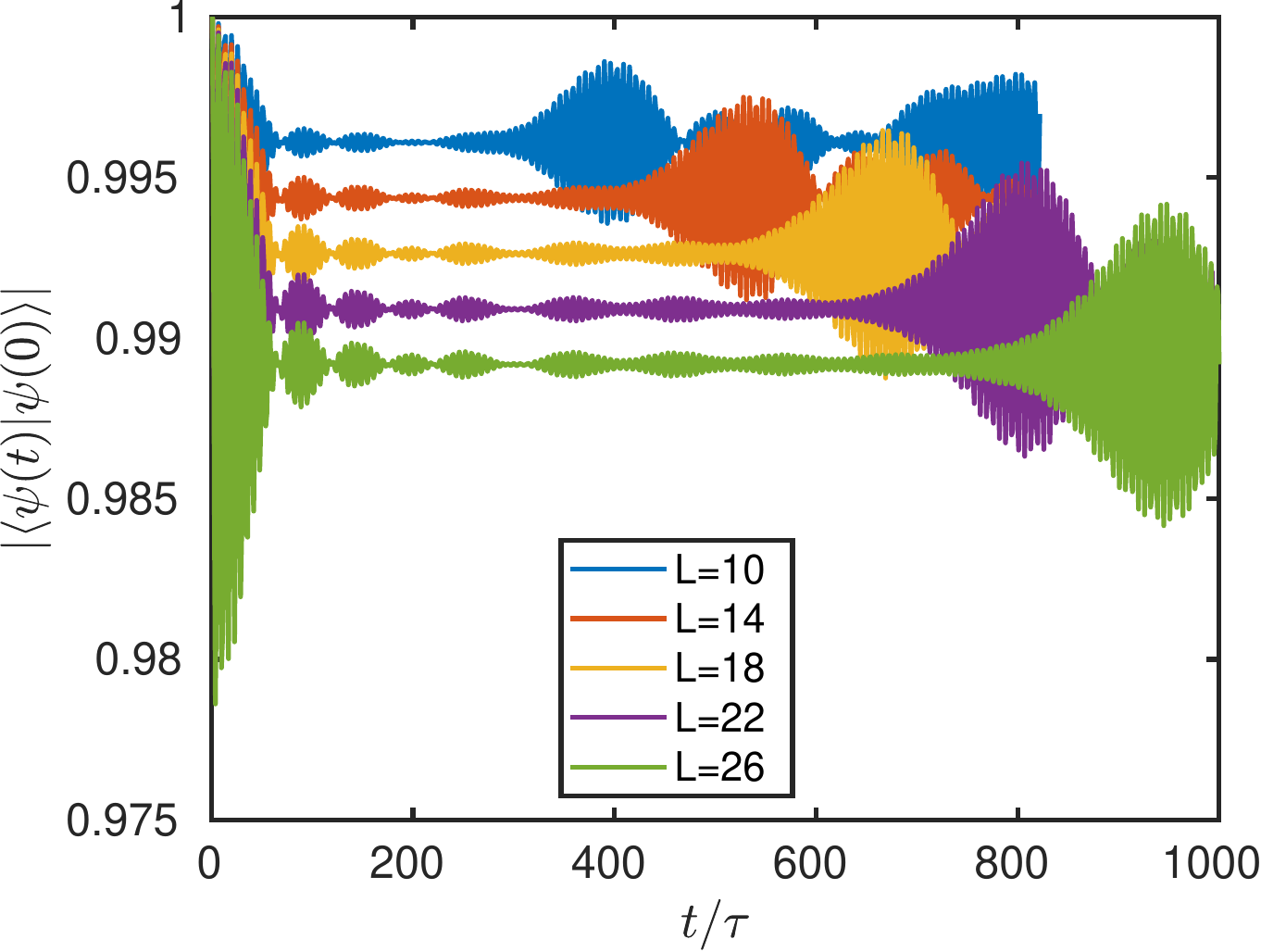}\\
   	\includegraphics[width=7cm]{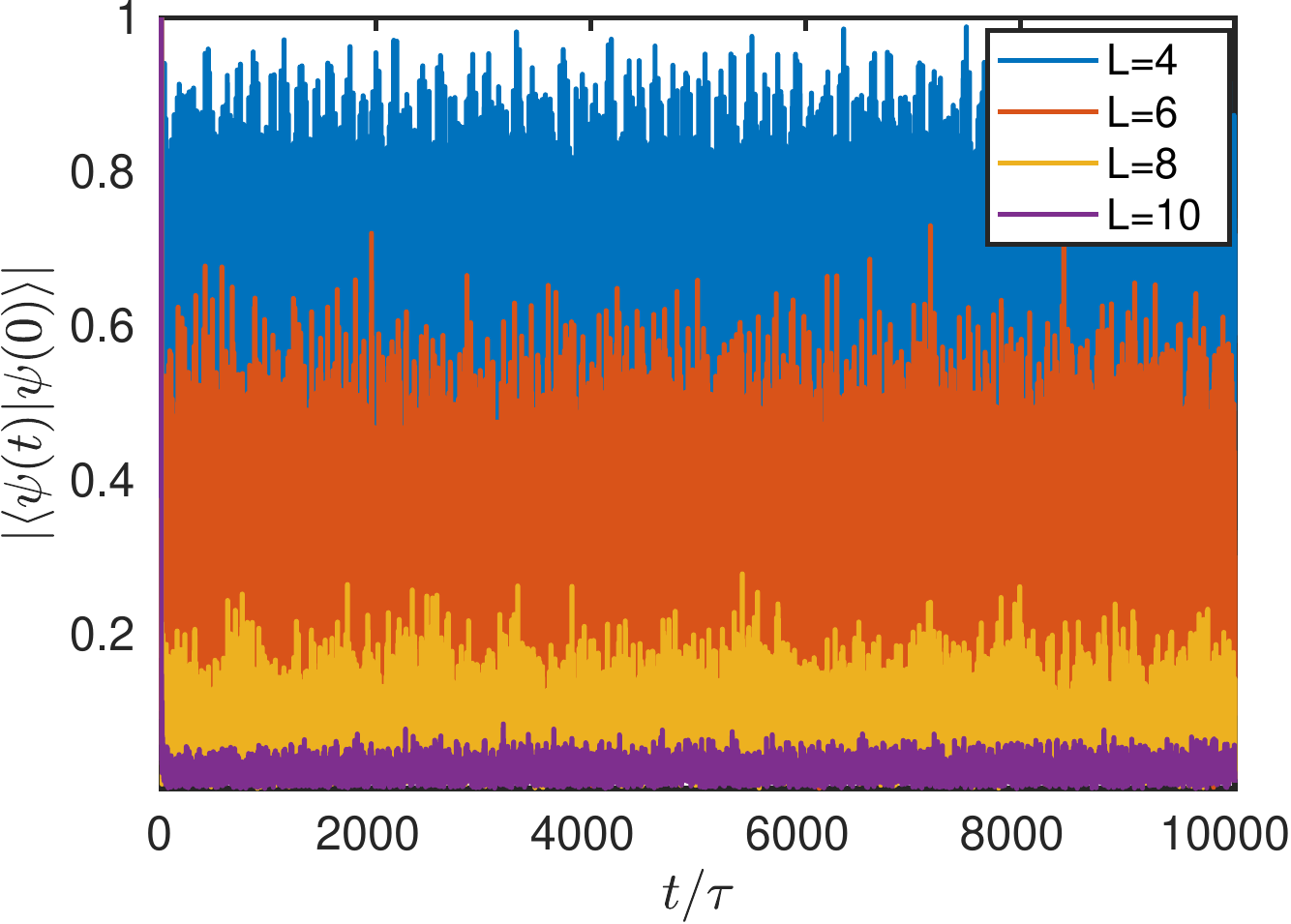}\\
   \end{tabular}
 \end{center}
 \caption{Overlap dynamics in a MBDL case (upper panel -- $J=0.01$, tDMRG with open boundary conditions) and a thermalizing case (lower panel -- $J=0.5$, full exact diagonalization).}
 \label{nonlinari:fig}
\end{figure}

\begin{figure}
	\begin{center}
		\begin{tabular}{c}
			\includegraphics[width=7cm]{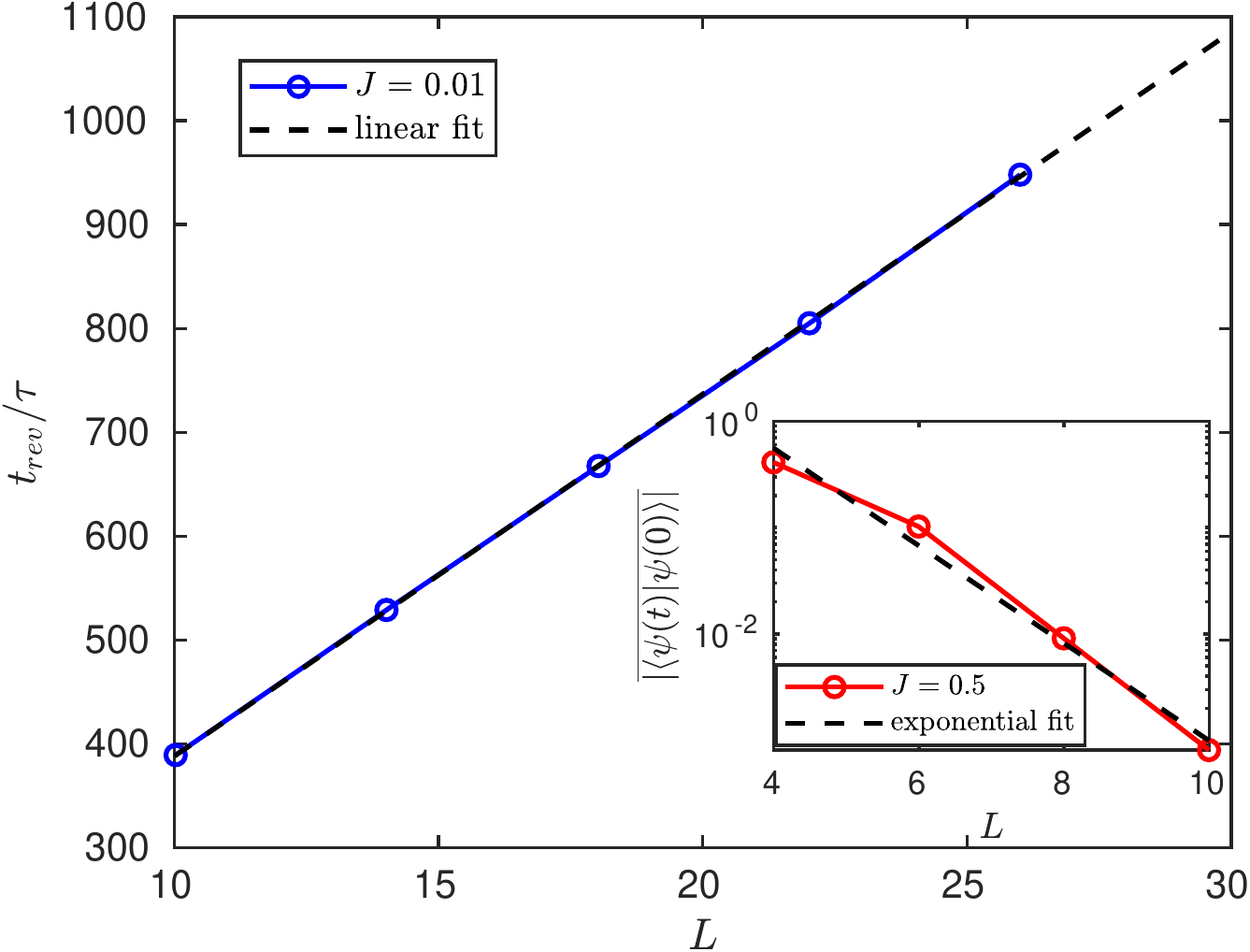}\\
		\end{tabular}
	\end{center}
	\caption{(Main panel) Scaling of the revival time $t_{rev}$ with the system size in a MBDL case ($J=0.01$). (Inset) {Time-averaged overlap versus the system size} in a thermalizing case ($J=0.5$). }
	\label{revival_scaling:fig}
\end{figure}

%
Another piece of evidence comes from the behaviour in time of the overlap with the initial state $\left|\bra{\psi(0)}\left.\psi(t)\right\rangle\right|^2$. There is a strong difference between the dynamically-localized case (upper panel of Fig.~\ref{nonlinari:fig}) and the thermalizing case (lower panel of Fig.~\ref{nonlinari:fig}). In the former case there is a very clear and strong finite-size revival at a time linearly increasing with the system size. (The revival time is the time of the second maximum and we plot it versus $L$ in the main panel of Fig.~\ref{revival_scaling:fig}). This revival is observed also in the dynamics of $\epsilon(t)$ (see examples in Fig.~\ref{eps_vs_L:fig}) and is strictly reminiscent of similar revivals observed in driven integrable systems. Moreover, the overlap stays near to 1, marking the fact that the state of the system is localized in the Hilbert space and does not move so much from its initial value. In the thermalizing case, in contrast, the overlap decays to a value exponentially small in the system size and never rises up. {Although system sizes are not very large and no definitive conclusion is possible, the time-averaged overlap shows a behaviour consistent with an exponential decay with the system size (inset of Fig.~\ref{revival_scaling:fig})}: The system is very delocalized in the Hilbert space (whose dimension is exponentially large in the system size). This difference of behaviour is strongly reminiscent of the Loschmidt echo which Peres found to show similarly different properties in time in integrable and chaotic few-degree-of-freedom cases~\cite{Peres_PRA84} (our definition of the overlap is not very different from the Loschmidt echo). In Appendix~\ref{02:app} we show that, with a different initial state, the overlap behaves in the same way in each case, remarking the robustness of the many-body dynamical localization to the initial state. 

\subsection{Entanglement entropy} \label{entanglement:sec}
%
Our numerical evidences show the existence of a non-thermal behaviour which persists as the system size is increased and is very different from the thermalizing one. 
In what is this phenomenon different from  many-body localization (MBL)? In order to address this question we discuss in the following the behaviour of the entanglement entropy Eq.~\eqref{entropy:eqn}. A smoking gun of the many-body localization is just the behaviour of this entropy. In the time domain it increases logarithmically~\cite{Montangero,Prosen_PRB08,Moore_PRL12,Serbyn_PRL13}, at variance with clean-integrable~\cite{Alba_PNAS,Alba_SciPostPhys_17,Cala_linea} and thermal~\cite{Singh_2016} systems where the entanglement entropy increases linearly in time (up to an extensive saturation value). Looking at the single eigenstates, the entanglement entropy in MBL obeys area law with strong fluctuations (see for instance~\cite{Pollmann_PRL14}) while in the thermalizing case it obeys volume law~\cite{PhysRevB.91.081103} and the fluctuations are small, because all the eigenstates of the dynamics are locally thermal.

All these findings are true for autonomous spin chains, but they have also been extended to driven spin chains~\cite{Lazarides_PRE14,PONTE_2015,ABANIN20161,PhysRevLett.114.140401}, to autonomous systems with a finite-dimensional local Hilbert space (as clock models~\cite{Federica_PRB19}) and to the static Bose-Hubbard chain with disorder, from a theoretical~\cite{Delande_APP17,Sierant1,Sierant2,Glen} and an experimental~\cite{Lukin256} point of view.  

\begin{figure}
 \centering
\includegraphics[width=70mm]{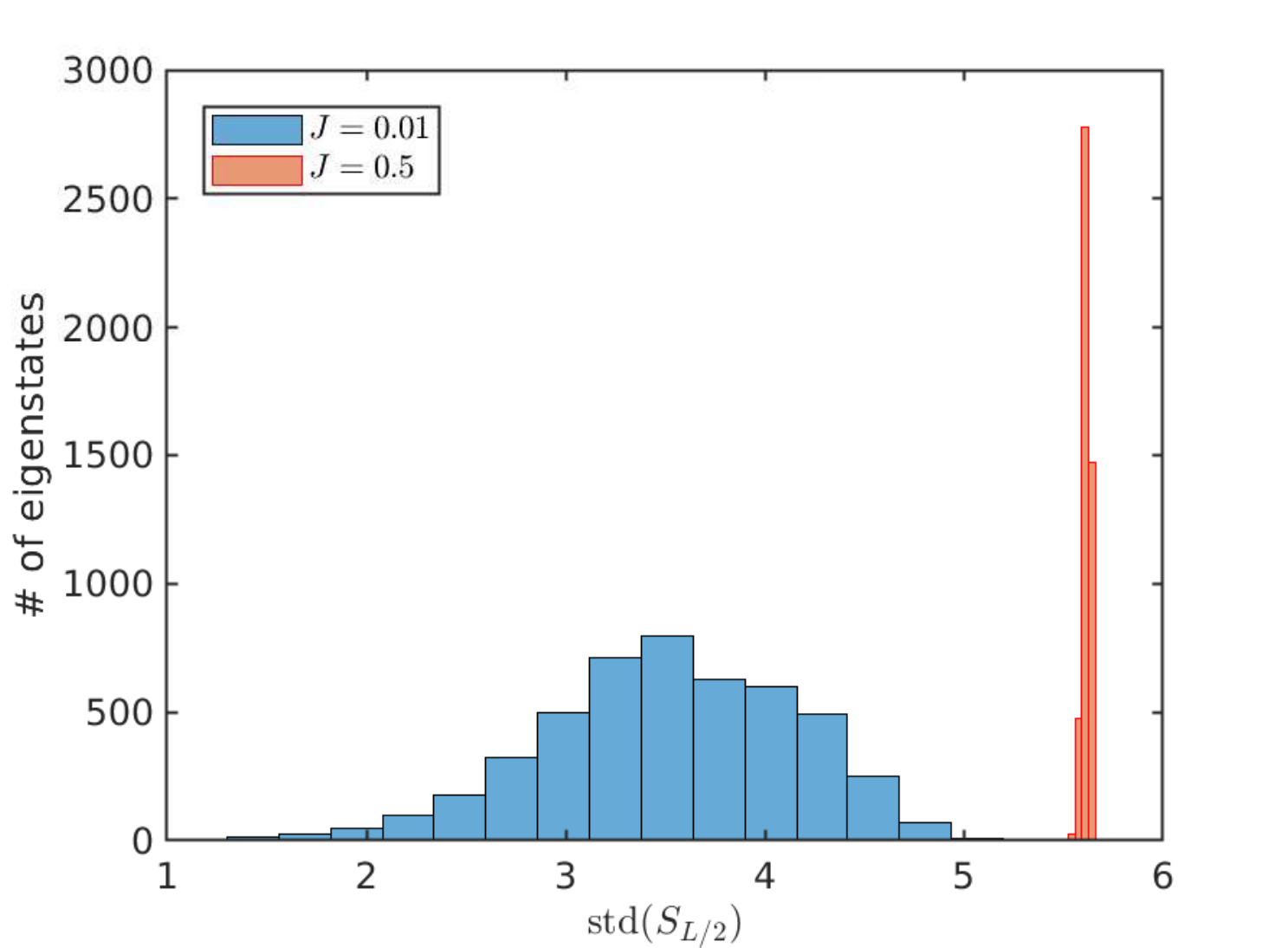}
\caption{Distribution of $S_{L/2}^\alpha$ for different values of $J$ and $L=10$.}
\label{pdfS:fig}
\end{figure}

We first study the difference between MBDL and thermalization by considering the half-chain entanglement entropy of the Floquet states. For each $\ket{\phi_\alpha}$ the corresponding entanglement entropy $S_{L/2}^\alpha$ is obtained through Eq.~\eqref{entropy:eqn}. First of all we consider the qualitative appearance of the distributions as we can see in Fig.~\ref{pdfS:fig}. Here we can clearly see that the distribution of the fully thermalizing case ($J=0.5$) is very peaked around the $T=\infty$ value (as appropriate for ETH) while the one for the dynamically localized case ($J=0.01$) is very {broad}. 

\begin{figure}
 \centering
\includegraphics[width=70mm]{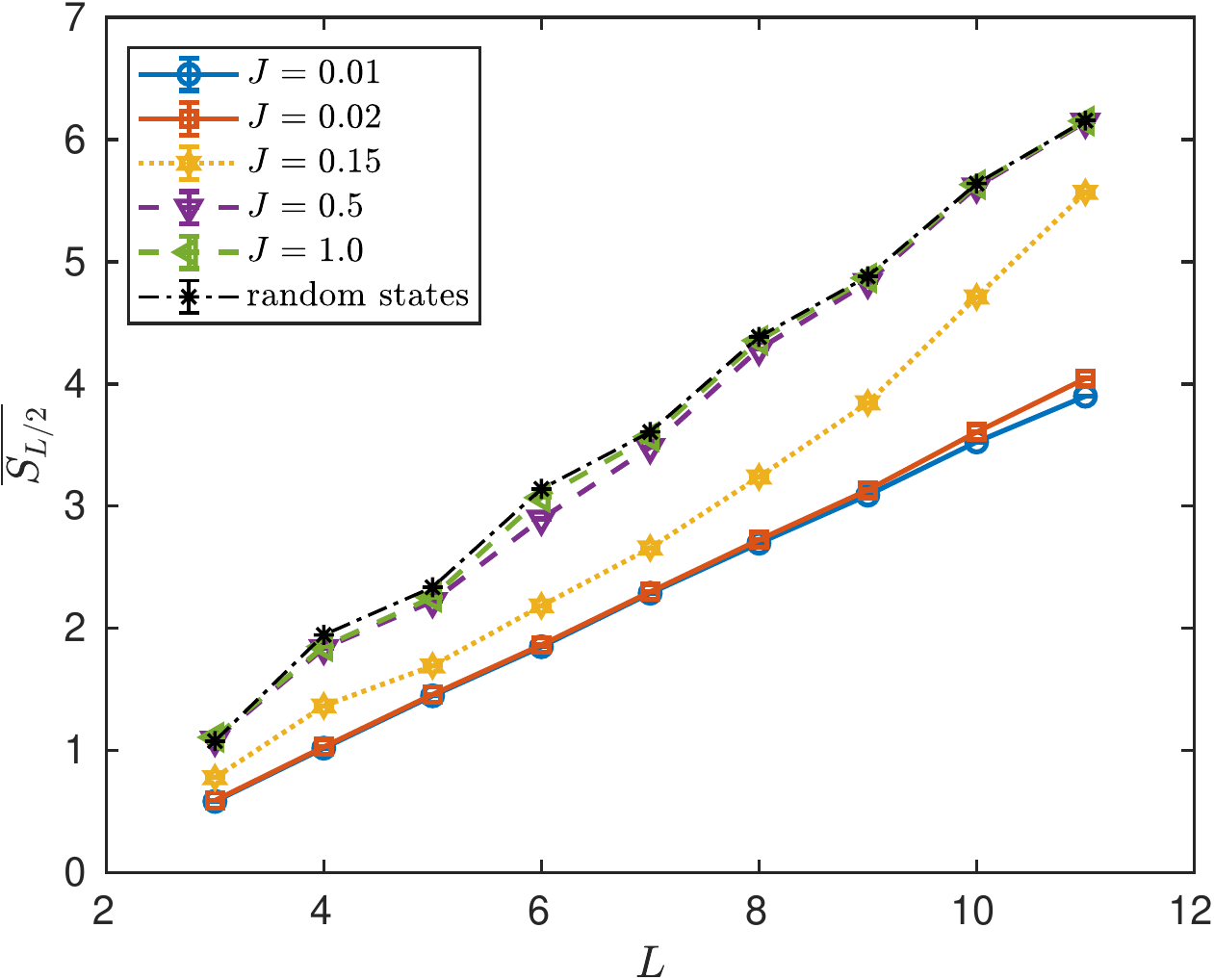}
\caption{Floquet-state average $\mean{S_{L/2}}$ versus $L$ for different values of $J$. {Notice the overall linear scaling which occurs with a significantly larger angular coefficient deep in the thermalizing phase ($J=0.5,\,1$) with respect to the MBDL phase ($J=0.01,\,0.02$). Deeply in the thermalizing phase the entanglement entropy is essentially coinciding with the corresponding value for a fully random state (the curve in figure is averaged over $N_{\rm real}=10000$ realizations of the randomness for $L\leq10$, $N_{\rm real}=500$ for $L=11$). For $J=0.15$ the thermal behaviour appears for larger system sizes.}}
\label{meanS:fig}
\end{figure}

{It is indeed important to do a scaling to clearly distinguish the two conditions. We first consider the scaling of $\mean{S_{L/2}}$, the average of the distribution (Fig.~\ref{meanS:fig})~\cite{Note2}. We consider two values of $J$ deep in the MBDL phase ($J=0.01$ and $J=0.02$), two values deep in the thermalizing phase ($J=0.5$, $J=1.0$) and one still thermalizing as suggested by  Fig.~\ref{exp_log_scaling:fig} but with a smaller value of $J$ ($J=0.15$). We see a linear scaling with the system size, both deep in the thermalizing phase and in the MBDL phases. So, on average the Floquet states obey volume law in both conditions. We compare these results with the entanglement entropy of a fully random state averaged over $N_{\rm real}>500$ randomness realizations, in the same spirit of the Page's analysis~\cite{page}. We see that the results of the deeply thermalizing cases are essentially equal to the fully random case. Therefore, thermalization coincides with the Floquet states being fully random objects, in agreement with the assumptions of ETH. Also the case $J=0.15$, after an interval of slower increase, bends towards the fully-random situation. In the dynamically localized case the linear increase is clearly different from the thermal one because it has a smaller angular coefficient and there is not the slight parity effect which appears there~\cite{Note3}. The linear increase in dynamical localization is very clear and the difference from the thermalizing case very patent, but the sizes we can access are limited and we cannot exclude that for some $L>11$ the curves start bending towards the thermal case, as it happens for $J=0.15$.}

\begin{figure}
 \centering
\includegraphics[width=70mm]{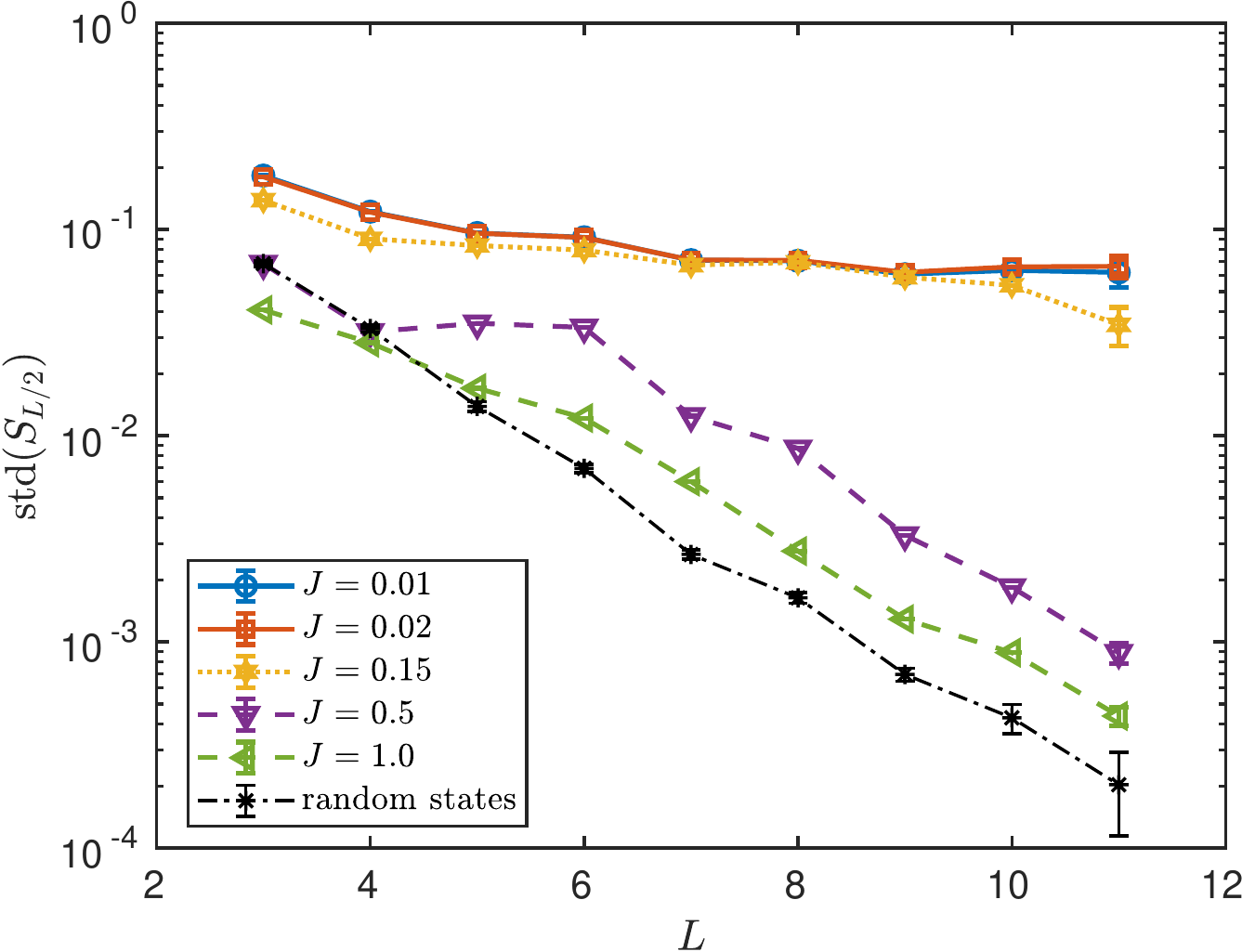}
\caption{Floquet-state standard deviation\; $\operatorname{std}(S_{L/2})/L$ versus $L$ for different values of $J$. {Notice the exponentially-fast convergence to 0 deep in the thermalizing phase ($J=0.5,\,1$) which is very similar to the behaviour obtained with fully random states (the random-state standard deviation is evaluated with the same number of realizations as in Fig.~\ref{meanS:fig}). On the opposite, in the MBDL cases ($J=0.01,\,0.02$) there is convergence to a finite value.}}
\label{stdS:fig}
\end{figure}

{In order to better perceive the difference between the two phases, we have to consider the scaling of\; $\operatorname{std}(S_{L/2})/L$, the standard deviation of $S_{L/2}^\alpha$ divided by $L$. We need this normalization because we want to consider the relative fluctuations with respect to an average which linearly scales with $L$. We perform this scaling in Fig.~\ref{stdS:fig} and we see a difference between the two phases as strong as the one observed in Fig.~\ref{exp_log_scaling:fig}. Deep in the thermal phase the relative fluctuations scale to 0 with increasing $L$ (and the scaling is clearly exponential), and this scaling closely mirrors the one of the standard deviation evaluated over a fully random state distribution. This is still in agreement with ETH and typicality. On the opposite, in the dynamically-localized case they attain a finite value. So, the entanglement distributions are different in the two cases also going towards the large-size limit.} 

\begin{figure}
 \centering
 \begin{tabular}{c}
 	\includegraphics[width=70mm]{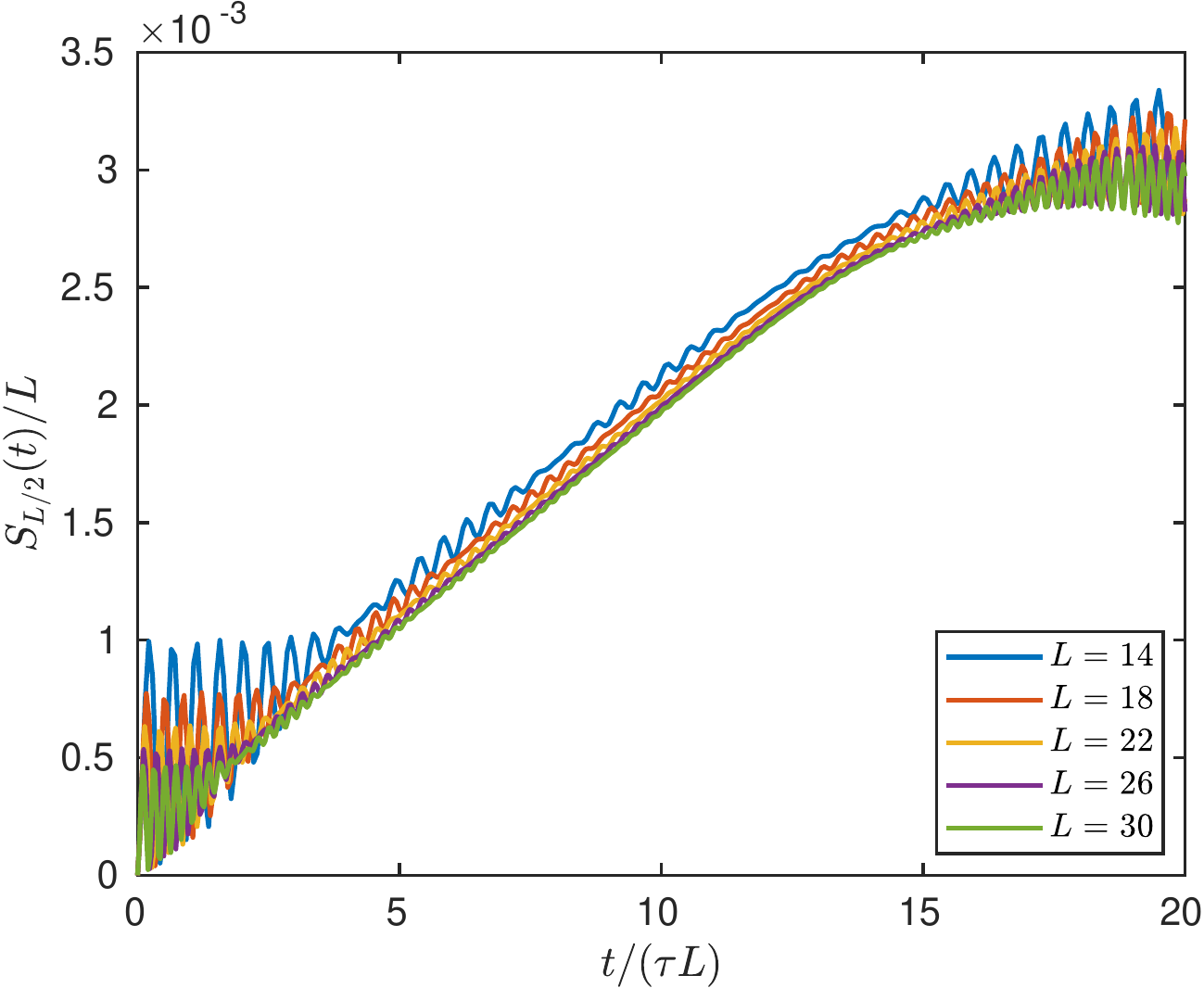}
 \end{tabular}
\caption{Rescaled half-chain entropy $S_{L/2}(t)/L$ as a function of $t/(\tau L)$ in a clean dynamically localized case ($J=0.01$). The entropy increases linearly up to a time $\sim t_{rev}/2$. Calculations performed using tDMRG and open boundary conditions.}
\label{entrop_ene:fig}
\end{figure}

\begin{figure}
 \centering
 \begin{tabular}{c}
 	\includegraphics[width=70mm]{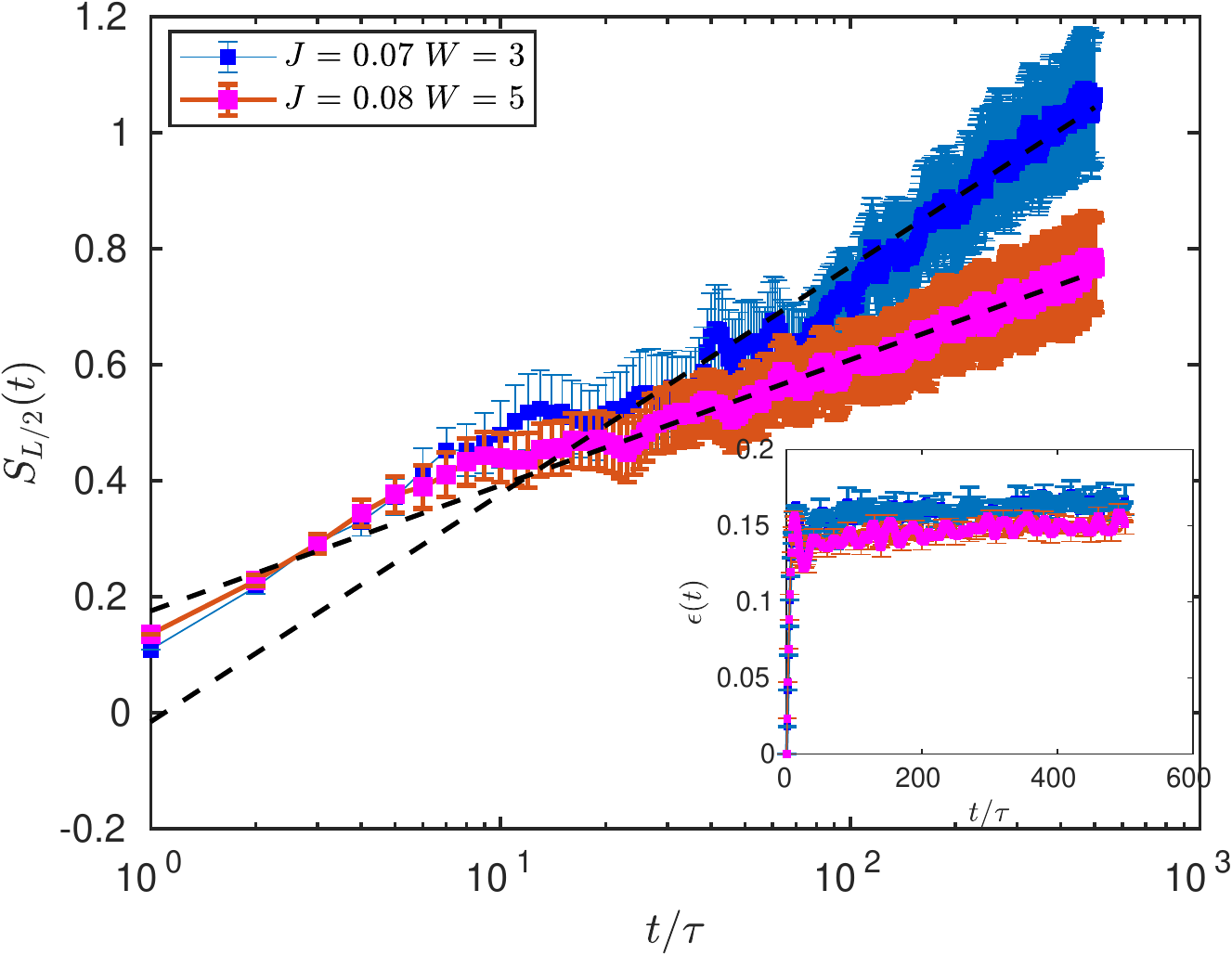}\\
 \end{tabular}
\caption{(Main figure) When a strong enough disorder is added, the half-chain entropy $S_{L/2}(t)$ increases much more slowly, in a way compatible with a logarithmic growth, as characteristic of the MBL phase. (Inset) In the MBL phase, also the energy absorption is hindered; we see that $\epsilon(t)$ [Eq.~\eqref{epsilont:eqn}] attains a condition where it oscillates around a value much smaller than 1. The errorbars both in $S_{L/2}(t)$ and $\epsilon(t)$ are evaluated as root mean square fluctuations over the $N_{\rm rand}$ disorder realizations.
(Other numerical parameters $L=20$, $N_{\rm real}=20$ for the blue $W=3$ curves and $N_{\rm real}=50$ for the red $W=5$ ones. Calculations performed using tDMRG and open boundary conditions.) 
}
\label{entrop_ene_MBL:fig}
\end{figure}

Now we move to study the difference between MBDL and
MBL by studying the behaviour in time of the half-chain entanglement entropy $S_{L/2}(t)$ for $t=n\tau^-$ (we put $\ket{\psi(n\tau^-)}$ in Eq.~\eqref{entropy:eqn}). In the MBDL case we see that it increases linearly in time until it saturates to an asymptotic value linear in the system size and much smaller than the thermal one (see Fig.~\ref{entrop_ene:fig}). This behaviour is very similar to the case of integrable driven systems~\cite{Russomanno_2016}. We can see, moreover, that the half-chain entanglement entropy obeys a finite-size scaling law $S_{L/2}(t)\sim L f(t/L)$ with $\lim_{x\to\infty} f(x)$ finite and order 1. This linear increase behaviour of the entropy is deeply different from the MBL one, where the entanglement entropy logarithmically increases in time. We can see the difference also in our system by adding disorder and inducing MBL. We put disorder in the form of a disordered chemical potential (as in~\cite{Glen,Lukin256,Sierant2}); the new Hamiltonian is
\begin{equation}
 \hat{H}_{\rm dis}(t) = \hat{H}(t) + \sum_j\mu_j\hat{n}_j\,,
\end{equation}
where $H(t)$ is given by the formula in Eq.~\eqref{Hamour:eqn} and $\mu_j$ is a random variable uniformly distributed in the interval $[-W,W]$. We consider $S_{L/2}(t)$ averaged over $N_{\rm real}$ disorder realizations. When the disorder amplitude $W$ is large enough, we find logarithmic increase of entanglement entropy (main panel of Fig.~\ref{entrop_ene_MBL:fig}), at variance with the linear increase of the MBDL case. Although the behaviour of the entanglement entropy is very different, the behaviour of $V(t)$ is non-thermalizing in both cases. We show this fact in the inset of Fig.~\ref{entrop_ene_MBL:fig} where we plot $\epsilon(t)$ averaged over the $N_{\rm real}$ disorder realizations and we see that after a transient it oscillates slightly and irregularly around a value much smaller than one, as appropriate for the absence of $T=\infty$ thermalization. So, it is very important to consider the behaviour of entanglement in order to distinguish MBL from many-body dynamical localization.

%
%
%
\section{Floquet states and localization in the Hilbert space} \label{fock:sec}
Here we look closer at the Floquet states and their localization properties in the Hilbert space.
We have already shown in Sec.~\ref{evidences:sec} that the expectation value of $\hat V$ over the Floquet states $\ket{\phi_\alpha}$ is characterized by a very large distribution in the MBDL phase. Here, instead we study what are the consequences of MBDL on the structure of each Floquet state. In particular, we will show that the {Floquet states are confined in subsectors of the Hilbert space quasi-degenerate in the undriven energy $\hat{V}$ (localization across quasi-degenerate multiplet subspaces). At the same time, Floquet states are largely delocalized inside these Hilbert space subsectors.}

\begin{figure}
 \begin{center}
   \begin{tabular}{c}
    \includegraphics[width=7cm]{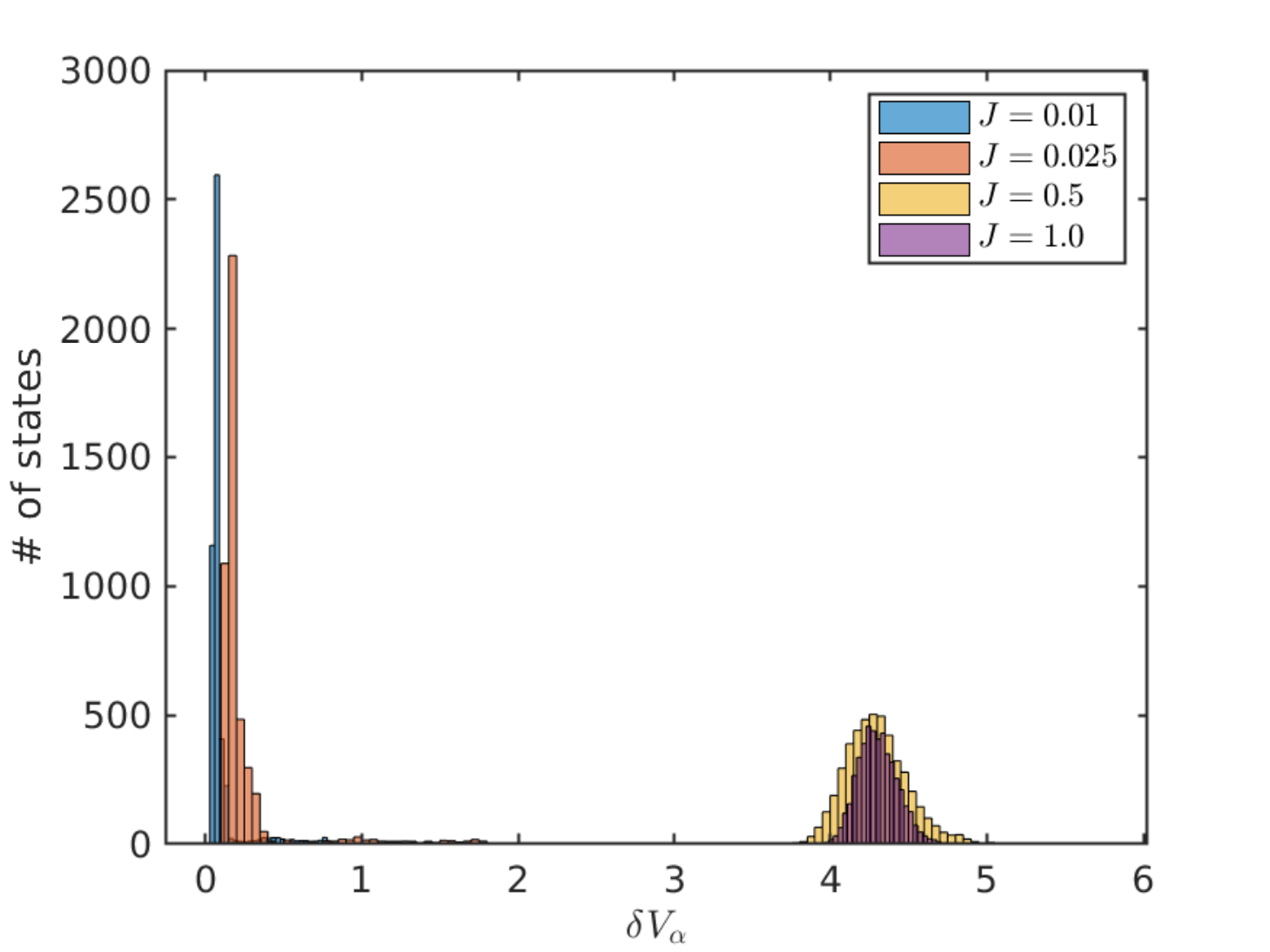}\\
   \end{tabular}
 \end{center}
 \caption{Distribution of $\delta V_\alpha$ for different values of $J$ and $L = 10$. Notice that the in the ergodic case the distribution is peaked at a value much greater than in the MBDL case.}
 \label{VdV:fig}
\end{figure}

\begin{figure}
	\begin{center}
		\begin{tabular}{c}
			\includegraphics[width=7cm]{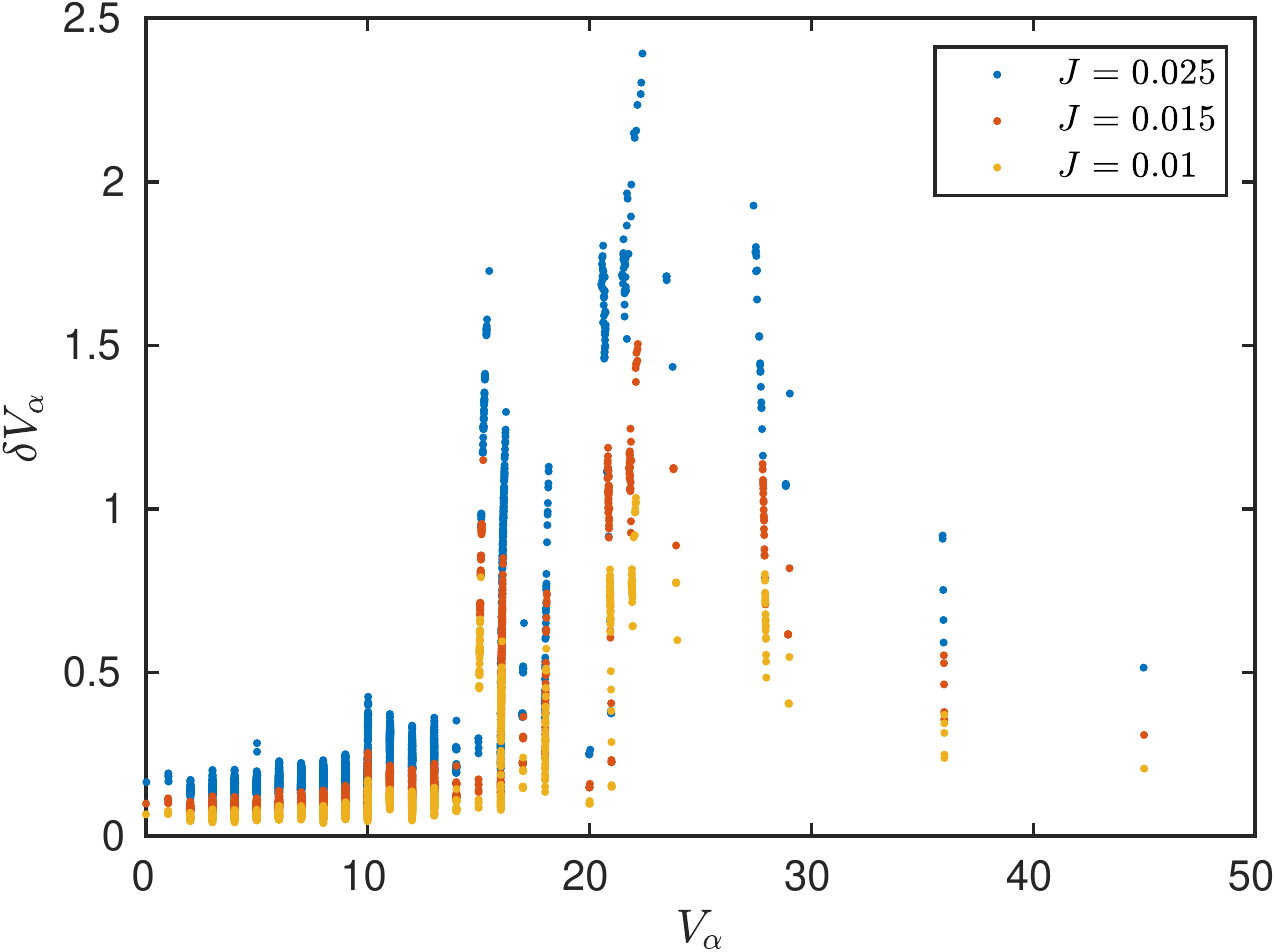}\\
			\includegraphics[width=7cm]{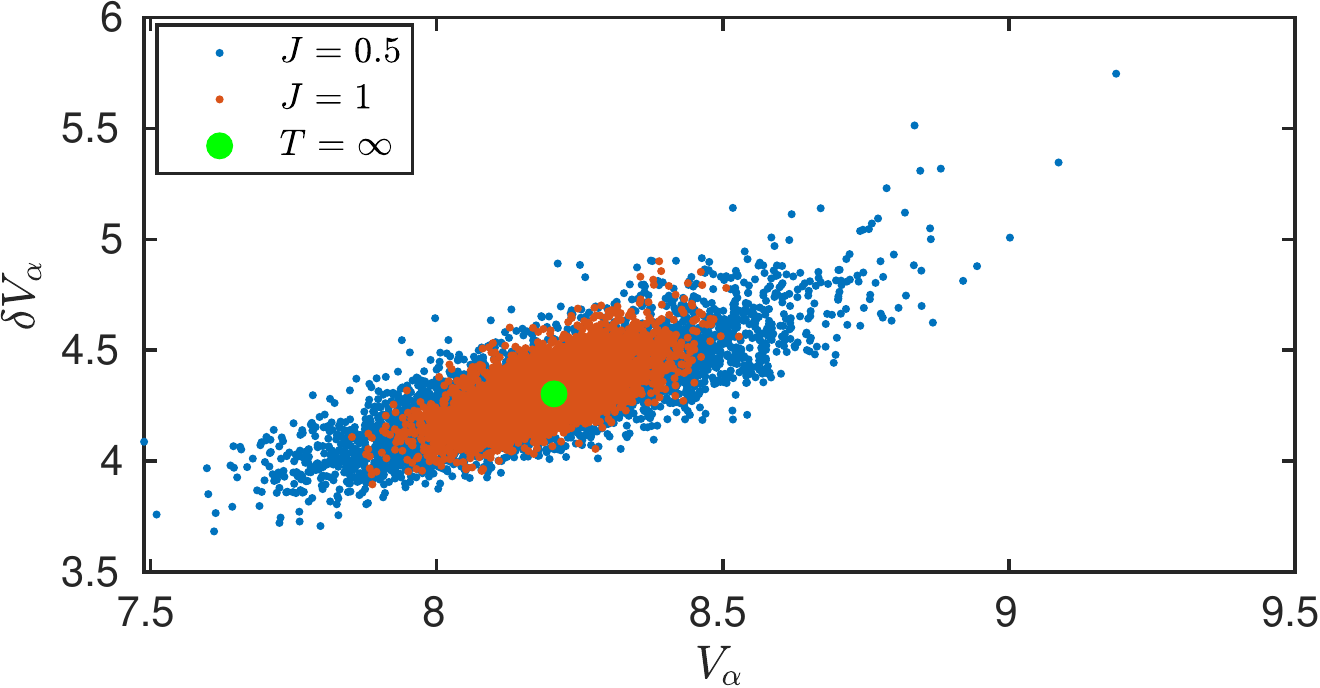}
		\end{tabular}
	\end{center}
	\caption{$\delta V_\alpha$ versus $V_\alpha$ for all the Floquet states $\ket{\phi_\alpha}$. Notice the difference between the dynamically localized regime (upper panel) and the thermalizing regime where ETH is fully developed (lower panel). Numerical parameters: $L=10$.}
	\label{VdV_joint_distribution:fig}
\end{figure}

In order to disentangle the localization across quasi-degenerate subspaces from other properties of a Floquet state, we define the undriven-energy fluctuation
\begin{equation} \label{delvalpha:eqn}
\delta V_\alpha\equiv\left(\Braket{\phi_\alpha|\hat{V}^2|\phi_\alpha}-\Braket{\phi_\alpha|\hat{V}|\phi_\alpha}^2\right)^{1/2}\,.
\end{equation}
%
We focus on the two limiting cases, deep in the MBDL and thermalizing phase. The distributions of $\delta V_\alpha$ are reported in Fig.~\ref{VdV:fig}. In both cases the distributions appear to be rather peaked, however the typical $\delta V_\alpha$ in the thermalizing phase is much larger than the typical $\delta V_\alpha$ in the MBDL phase. 
The difference in the structure of the Floquet states is graphically well captured in the $\delta V_\alpha$ versus $V_\alpha$ plots of Fig.~\ref{VdV_joint_distribution:fig}. {In the ergodic case both $V_\alpha$ and $\delta V_\alpha$ have a quite narrow distribution around the $T=\infty$ ETH values, consistently with typicality (Fig.~\ref{VdV_joint_distribution:fig}, lower panel). On the opposite, in the MBDL case (Fig.~\ref{VdV_joint_distribution:fig}, upper panel) $\delta V_\alpha$ remains small and the range of $V_\alpha$ is large.~\footnote{We discuss in Appendix~\ref{hybrid:app} some features of a case with $J$ intermediate between these two conditions.} (The small $\delta V_\alpha$ value in the MBDL phase was already to be expected from Fig.~\ref{pdf:fig}.) In particular, the values of $V_\alpha$ are arranged in quasi-degenerate multiplets. The small-amplitude kicking induces mixing inside the degenerate subspaces of $\hat{V}$ but not across the subspaces (at least for the sizes we can consider).  In the MBDL case the $V_\alpha$ gather around certain values, that at these system sizes are close to the unperturbed eigenvalues of $\hat V$. However, we are honestly unable to tell if this clustering is a finite-size effect and at larger system sizes the $V_\alpha$ distribution will be still broad but smooth over the real axis.}

\begin{figure}
	\begin{center}
		\begin{tabular}{c}
			\includegraphics[width=7cm]{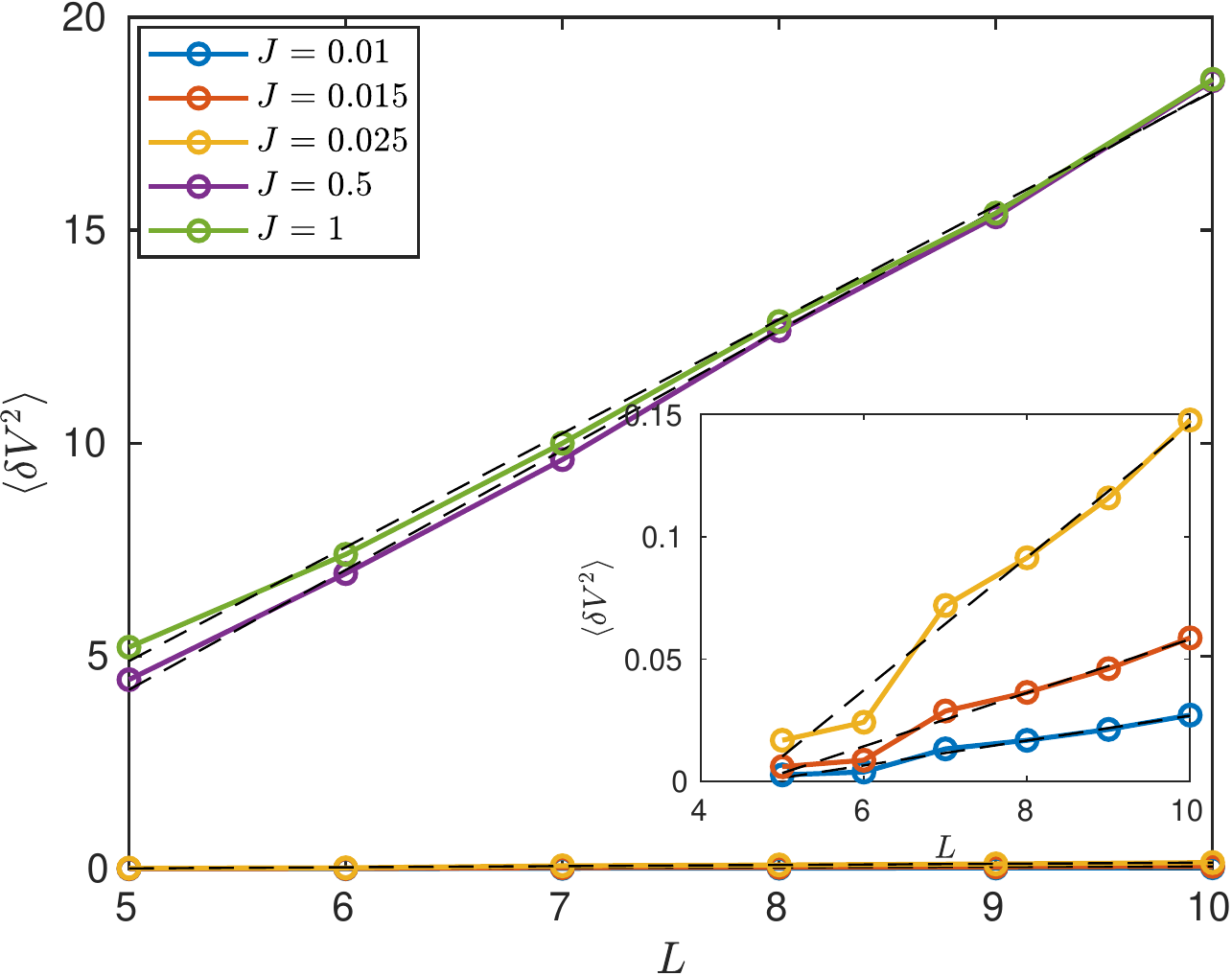}\\
		\end{tabular}
	\end{center}
	\caption{Scaling of $\langle \delta V^2 \rangle$ with $L$ for different $J$ values. The inset is a zoom in on the lower $J$ values. It appears that in both cases $\langle \delta V^2 \rangle\varpropto L$.}
	\label{dV_scaling:fig}
\end{figure}

{To better understand this point,} we have performed a scaling analysis on the width of the states $\delta V_\alpha$ in the two phases. We have focused on the average value of $\delta V_\alpha^2$ over all Floquet states: $\langle \delta V^2 \rangle$. Both in the MBDL and thermalizing phases {we find a linear scaling with the system size, $\langle \delta V^2 \rangle\varpropto L$. We show this scaling in Fig.~\ref{dV_scaling:fig} and we see that the energy width of a Floquet state in the MBDL phase is smaller than its counterpart in the thermalizing phase by at least { two orders of magnitude}.}

{Before discussing the physical meaning of this difference of order of magnitude, let us focus on the linear form of the scaling with $L$ of $\langle \delta V^2 \rangle$. This linear form was to be expected in both phases.} Indeed, imagine taking two chains ($A$ and $B$) of length $L$ without any interaction between them. The Floquet states of the two-chains system will just be a product of two Floquet states on the two subsystems $\ket{\Phi_{\alpha\beta}}_{AB} = \ket{\phi_\alpha}_A \otimes \ket{\phi_\beta}_B$. So the variance of $\hat{V}$ over $\ket{\Phi_{\alpha\beta}}$, $\Braket{\Phi_{\alpha\beta}|\hat{V}^2|\Phi_{\alpha\beta}}-\Braket{\Phi_{\alpha\beta}|\hat{V}|\Phi_{\alpha\beta}}^2$, can be computed to be $\delta V_\alpha^2 + \delta V_\beta^2$. So on average $\langle \delta V^2\rangle_{2L} = 2 \langle \delta V^2\rangle_{L}$ if there is no interaction between the two chains. If we now connect the two chains, we might expect that this further interaction, if anything, will spread the Floquet states even more in energy, so when we have interaction $\langle \delta V^2\rangle_{2L} \gtrsim 2 \langle \delta V^2\rangle_{L}$. In the MBDL phase, this interaction seems not to spread the Floquet states further, thus $\langle \delta V^2\rangle_{2L} = 2 \langle \delta V^2\rangle_{L}$ and we have a linear behaviour.
For what concerns the thermalizing phase, the linear scaling of $\delta V^2$ can instead be understood by assuming (as we checked in the lower panel of Fig.~\ref{VdV_joint_distribution:fig}) that it can be computed directly in the $T=\infty$ ensemble $\delta V_\alpha^2 \simeq \left({V}^2\right)_{T=\infty}-({V}_{T=\infty})^2$. This last quantity, as customary with fluctuations in thermal ensembles, scales proportionally to $L$.

{In conclusion, there is a regime where Floquet states are very narrow in energy and this corresponds to many-body dynamical localization. In the MBDL phase the Floquet states are localized across quasi-degenerate energy subspaces, at variance with the thermalizing phase, where $\langle \delta V^2 \rangle$ is two orders of magnitude bigger. Let us focus on the MBDL regime, in order to better understand here the localization properties of the Floquet states. We claim that they are localized across energy multiplet subspaces but at the same time they are substantially delocalized inside these subspaces. Delocalization can occur inside the degenerate eigenspaces of $\hat V$  which are highly dimensional objects (given a string of occupation $\nb$, each of its permutations will yield the same value of $V$).} 

{These eigenspaces are of course altered by the presence of the driving and, 
in order to discuss the question in a rigorous way, we need to estimate their dimension. For that purpose, we start introducing the energy distribution of a Floquet  state $\ket{\phi_\alpha}$. We define this distribution as the set of coefficients $p_v^\alpha = \braket{\phi_\alpha|\hat{\Pi}_v|\phi_\alpha}$, with $\hat{\Pi}_v$ being the projector on the eigenspace of $\hat V$ with eigenvalue $v$.
For each Floquet state $\ket{\phi_\alpha}$ we define the minimum possible IPR achievable by a normalized state $\ket{\psi}$ which shares the same energy distribution of $\ket{\phi_\alpha}$ ($p_v^\alpha$ for $v\in$ the eigenvalues of $\hat{V}$). We dub this last quantity IPR$_{\text{deloc}}$, which is more explicitly given by
\begin{equation}
	\mbox{IPR}_{\text{deloc}}^\alpha = \sum_v \frac{p_v^\alpha}{\mbox{dim}\left[\mbox{Ker}\left(\hat V - v\boldsymbol{1}\right)\right]}\,.
\end{equation}
%
{where the symbol $\mbox{Ker}$ means the kernel of the operator in the argument.
The $\mbox{IPR}_\alpha$ defined in Eq.~\eqref{IPR:eqn} attains the value $\mbox{IPR}_{\text{deloc}}^\alpha$ if the state $\ket{\phi_\alpha}$ is maximally delocalized in the subspace available to it, once the distribution $p_v^\alpha$ is fixed. Therefore, according to the general properties of IPR, $\mbox{IPR}_{\text{deloc}}^\alpha$ is an estimate for the inverse of the effective dimension of the Hilbert space available to $\ket{\phi_\alpha}$. (Deep in the MBDL phase, $\mbox{IPR}_{\text{deloc}}$ essentially coincides with the inverse dimension of the eigenspace of $\hat{V}$ where $\ket{\phi_\alpha}$ is located and for which $p_v^\alpha\simeq 1$.)}} 

\begin{figure}
 \centering
 \begin{tabular}{c}
   {\includegraphics[width=70mm]{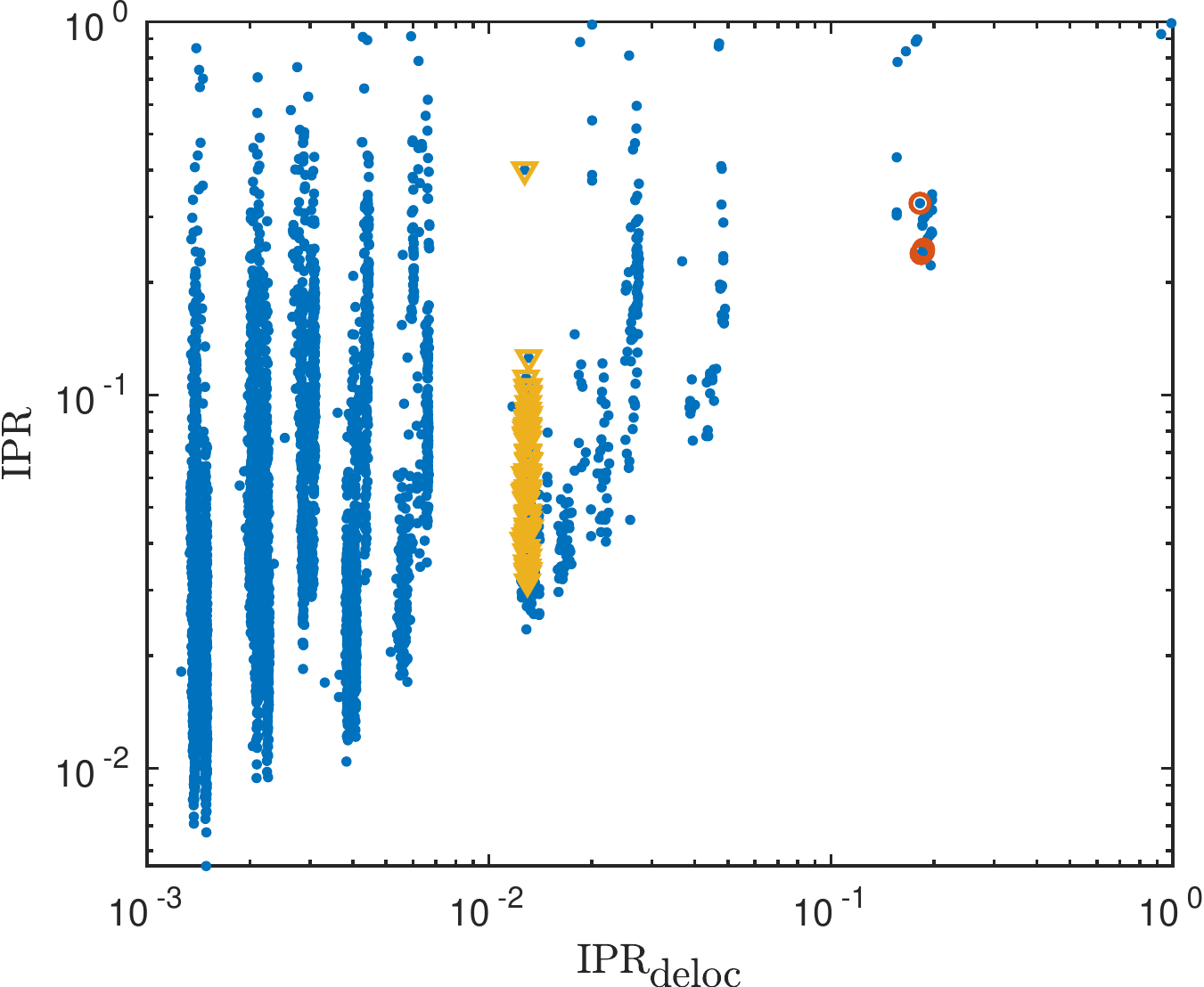}}
 \end{tabular}
\caption{IPR$_\alpha$ versus $\mbox{IPR}^\alpha_\text{deloc}$ for the different Floquet states $\ket{\phi_\alpha}$. We can notice a quite strong correlation between $\mbox{IPR}^\alpha_\text{deloc}$ and $\text{IPR}_\alpha$. The states with $V_\alpha\simeq1$ and $V_\alpha\simeq2$ are further denoted by a red circle and a yellow triangle, respectively. (Numerical parameters: $L=10$, $J=0.03$.)}
\label{fig:IPR_vs_IPR_deloc}
\end{figure}

{So, in order to understand the delocalization properties of each Floquet state $\ket{\phi_\alpha}$ we have to take the inverse of the effective Hilbert space dimension (estimated as $\mbox{IPR}_{\text{deloc}}^\alpha$) and compare it with the IPR$_\alpha$. Usually the comparison is done with the dimension of the full Hilbert space, but here there is the constraint of localization across the quasi-degenerate energy subspaces and we have to use the effective dimension.} In Fig.~\ref{fig:IPR_vs_IPR_deloc} we report the joint distribution of IPR$_\alpha$ and $\mbox{IPR}^\alpha_{\text{deloc}}$ in the MBDL phase.
We note that IPR distributions are quite {broad}. In particular, there are many Floquet states with small IPR that would seem to be delocalized. Furthermore, there seems to be a correlation between $\mbox{IPR}^\alpha_{\text{deloc}}$ and the IPR$_\alpha$ {(smaller values of $\mbox{IPR}^\alpha_{\text{deloc}}$ allow smaller values of IPR$_\alpha$), suggesting that Floquet states are mainly delocalized in the Hilbert space region available to them.}

\begin{figure}
 \centering
 \begin{tabular}{c}
   {\includegraphics[width=70mm]{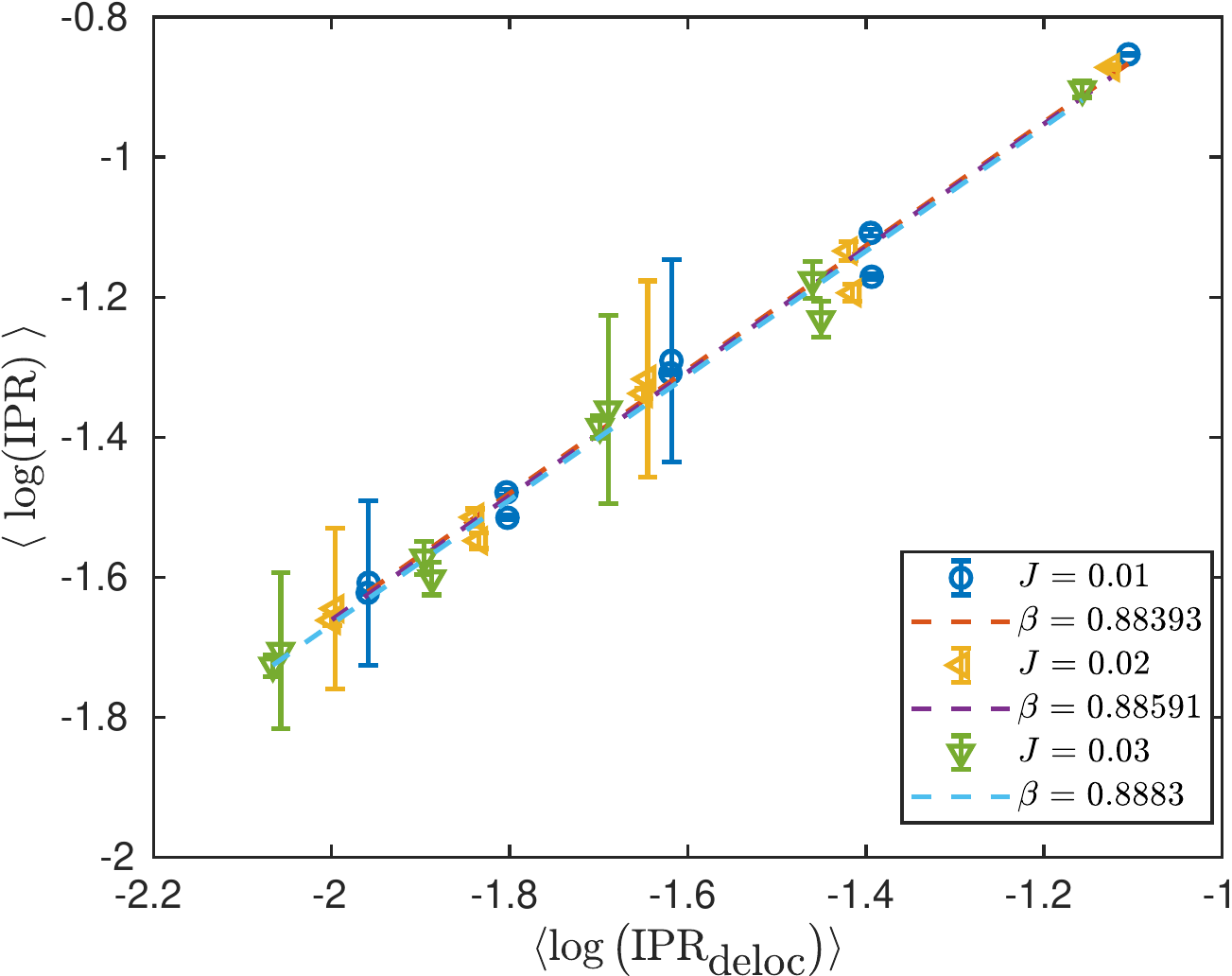}}\\
   {\includegraphics[width=70mm]{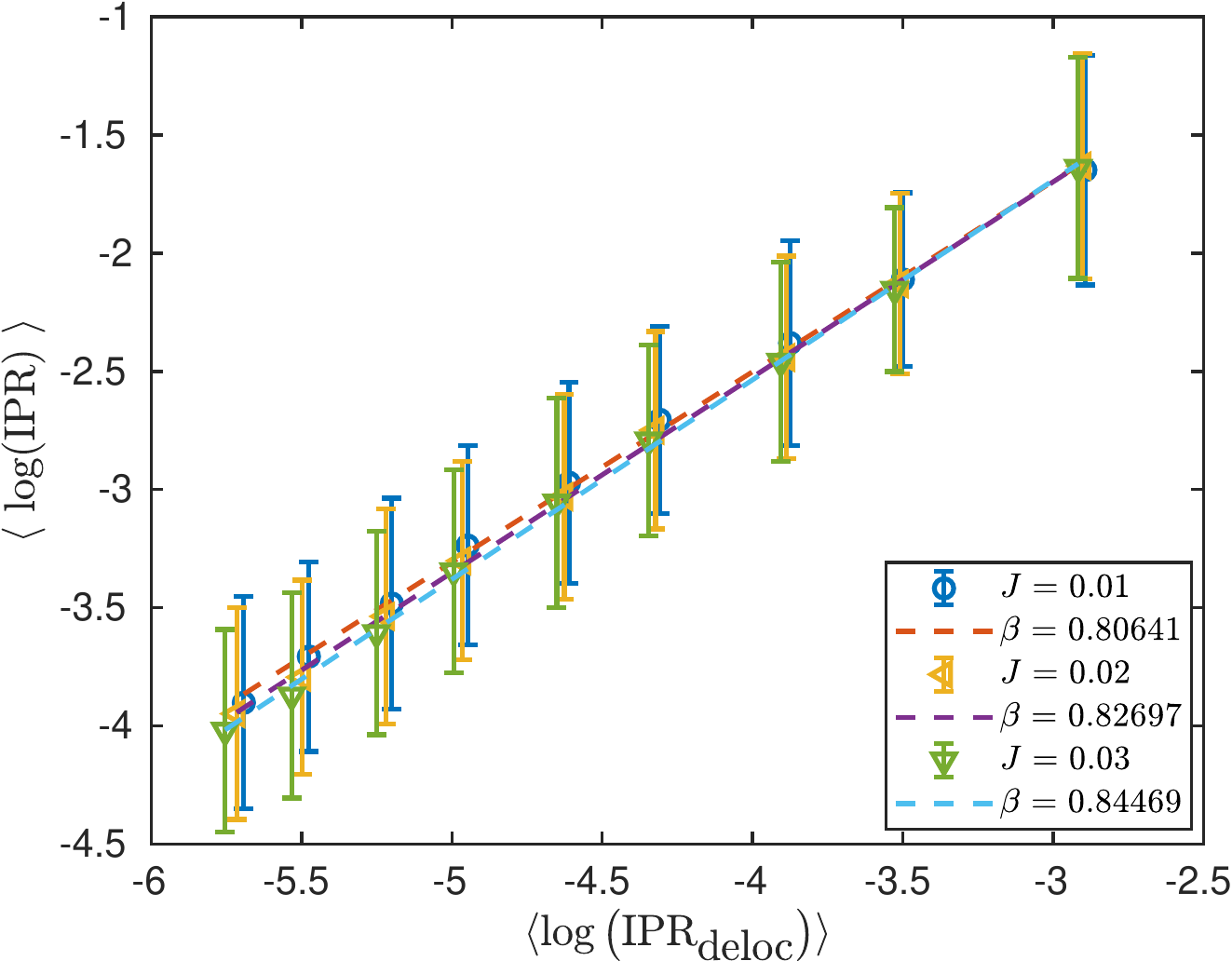}}
 \end{tabular}
\caption{Average $\log\text{IPR}$ against the average degeneracy $\log\text{IPR}_{\text{deloc}}$ defined in text. In the upper (lower) panel averages are performed over Floquet states with $V_\alpha = 1$ ($V_\alpha = 2$). In both cases a power law behaviour $\langle\log\text{IPR}\rangle\sim\gamma + \beta \langle \log\text{IPR}_\text{deloc} \rangle $ can be recognised. An errorbar on each point denotes the standard deviation of $\log$IPR over the multiplet subspace. Each point corresponds to a
value of $L$, from $L = 6$ (rightmost point) to $L = 14$ (leftmost point).}
\label{fig:IPR_IPR_uni_scaling}
\end{figure}

In order to obtain a more quantitative understanding of this delocalization, we focus on a set of Floquet states taken from a narrow $V_\alpha$ window ({corresponding to one of the quasi-degenerate subspaces of $\hat{V}$ introduced above}), so that we can expect them to have similar energy constraints. We study how both the IPR and the $\mbox{IPR}_{\text{deloc}}$ decrease as the system size is increased. In fact, if the states are delocalized, we would expect a scaling $\mbox{IPR}\sim\mbox{IPR}_{\text{deloc}}$, while if they are perfectly localized we might expect that the IPR will not scale at all as $\mbox{IPR}_{\text{deloc}}$ will decrease.
To perform this study, we have restricted ourselves to states with few excitations, which we can study up to $L=14$ employing a truncation scheme with $\Delta_T=3$. Namely we will focus on two sets, selected requiring that $V_\alpha\simeq 1$ and $V_\alpha\simeq 2$ respectively (highlighted in Fig.~\ref{fig:IPR_vs_IPR_deloc}).
We show in Fig.~\ref{fig:IPR_IPR_uni_scaling} that, within a given set, the logarithmic average of the $\mbox{IPR}$ of these states scales as a power law with the logarithmic average of $\mbox{IPR}_{\text{deloc}}$. More precisely, it appears that the scaling behavior is compatible with 
$\langle\log\left(\mbox{IPR}\right)\rangle = \gamma + \beta\langle\log\left(\mbox{IPR}_{\text{deloc}}\right)\rangle$
where $0<\beta<1$ and the angular brackets denote the average over the set. A positive scaling exponent $\beta$ signals that Floquet states are indeed delocalized in the portion of Hilbert they can access {(roughly, the $\hat V$ degenerate eigenspace they participate in)}.

Delocalization inside the $\hat V$ eigenspaces signals that excitations, while being localized in energy space, are delocalized in physical space and can travel along the chain. So, from one side there is energy localization connected to dynamical localization and from the other side there is spatial delocalization due to mixing of the Fock states inside the same $\hat V$ {multiplet} eigenspace. This is again in stark contrast both with thermalizing systems and MBL ones. Delocalization in space provides a framework for explaining the linear increase of the entanglement entropy. Indeed, in clean integrable cases, this linear increase is intimately connected to the delocalization in space and to the ballistic propagation of excitations~\cite{Alba_SciPostPhys_17}. From the other side, localization in energy space explains why the dynamics does not explore all the {Hilbert space, ergodicity is broken and the energy absorption is hindered.}

\section{Comparison with truncated Wigner approximation}
\begin{figure}
 \centering
 \begin{tabular}{c}
  {\includegraphics[width=70mm]{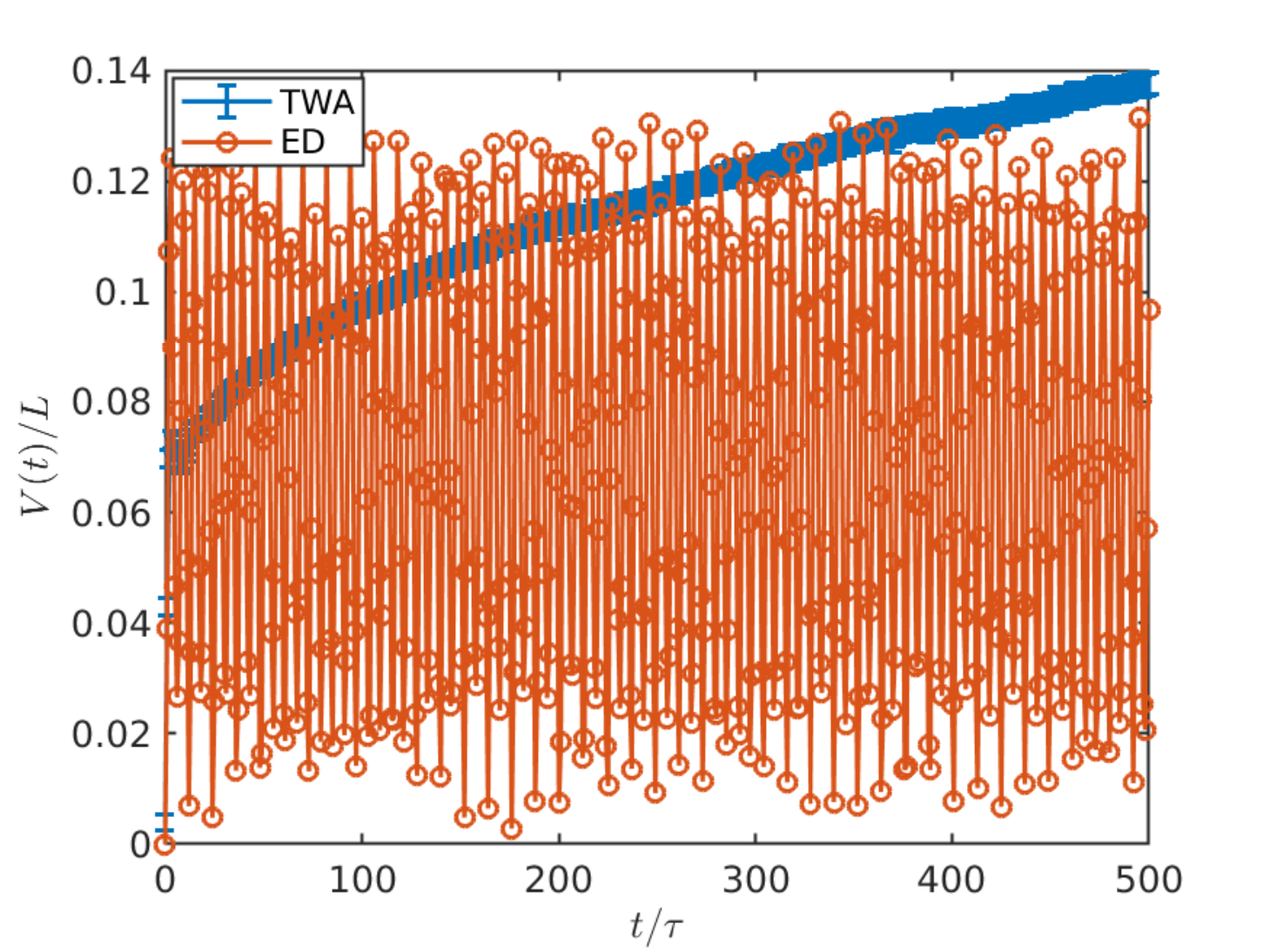}}\\
  {\includegraphics[width=70mm]{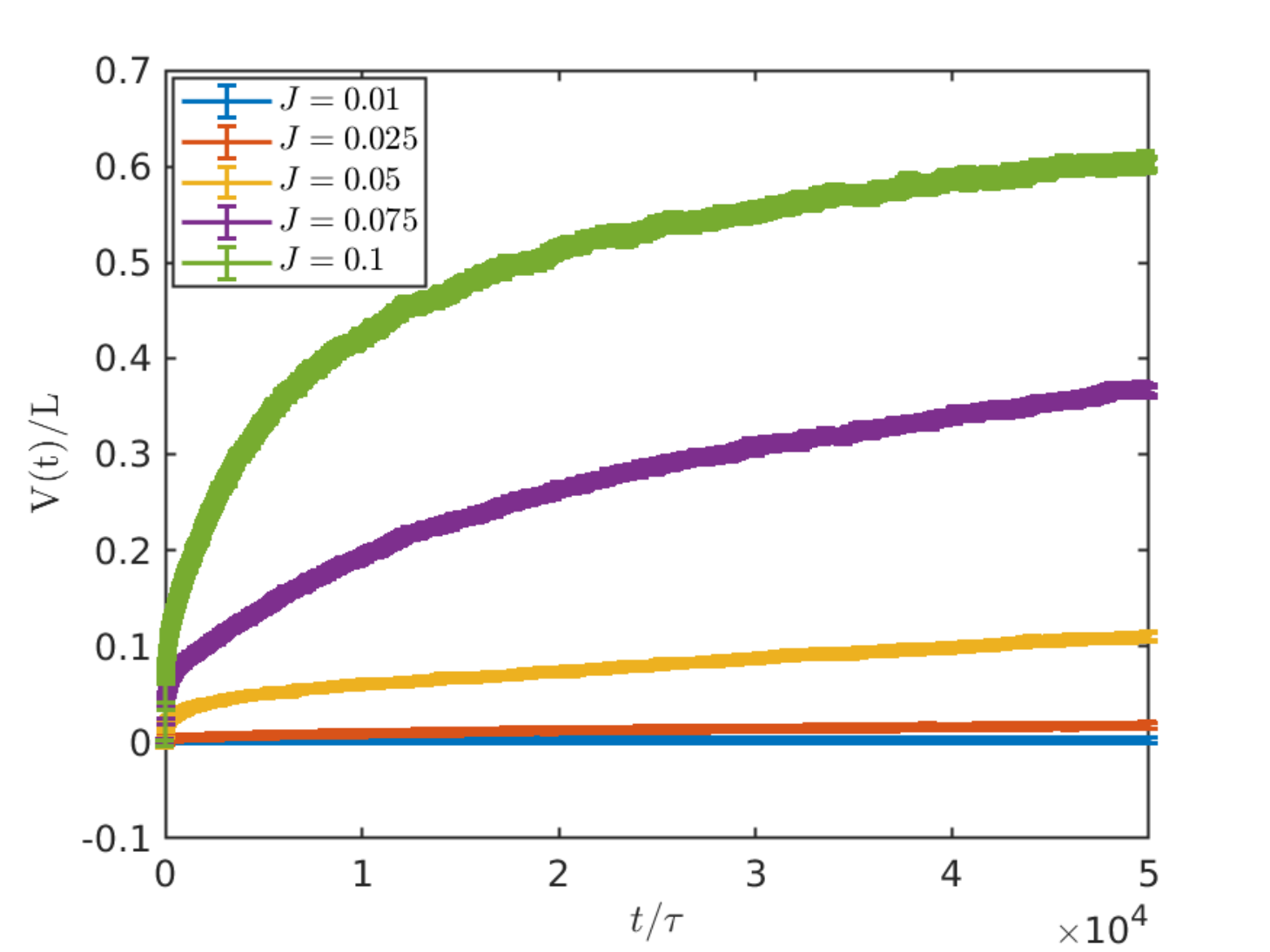}}
 \end{tabular}
\caption{Panel (a): for $L=5$ and $J=0.1$ the time evolution in the TWA and the exact one (ED) are compared. Panel (b): The time evolution for $L=5$ and various $J$. In both panels $\nu=1$.}
\label{fig:TWA}
\end{figure}

%
\label{sec:TWA}
Before concluding we would like to show a crucial point, that the many-body dynamical localization we are witnessing is a true quantum effect with no classical counterpart, exactly as the other forms of dynamical localization described in the literature~\cite{Boris:rotor,Casati:rotor,Notarnicola_PRB18}. In order to do that, we will show that dynamical localization disappears when the truncated Wigner approximation (TWA) is used to break quantum coherence.
This approximation has been described in great detail in Ref.~\cite{POLKOVNIKOV_TWA} and the underlying idea is to work in classical phase space employing Wigner quasi-distributions and Weyl symbols of operators. We are going to explain all the technical details in Appendix~\ref{TWA:app}, {here we just remark that this approximation amounts to put to zero the commutator between the operator $\hat{n}_j$ and the operator $\hat{\varphi}_j$ defined as $\hat{a}_j=\sqrt{\hat{n}}\nep^{i\hat{\varphi}_j}$. Because this commutator is actually $-i$, this is a {\it classical} approximation and amounts to break (after a small transient) quantum coherence in the evolution of the system.} We report in Fig.~\ref{fig:TWA} the time evolution from the uniform state obtained both through the TWA and exact diagonalization. In the exact quantum dynamics the energy absorbed reaches its asymptotic value after the first few kicks, but then it stops growing in time, while oscillating around its stationary value. %
On the opposite, the TWA dynamics is completely different. Also here, in the first few kicks the system absorbs the same amount of energy as in the quantum dynamics. However at longer times, the energy starts growing diffusively, {\it i.~e.} $V(t)/L = D t+{\rm const.}$, until at much greater times it will eventually reach the $T=\infty$ thermal value. 

\begin{figure}
 \centering
 \includegraphics[width=70mm]{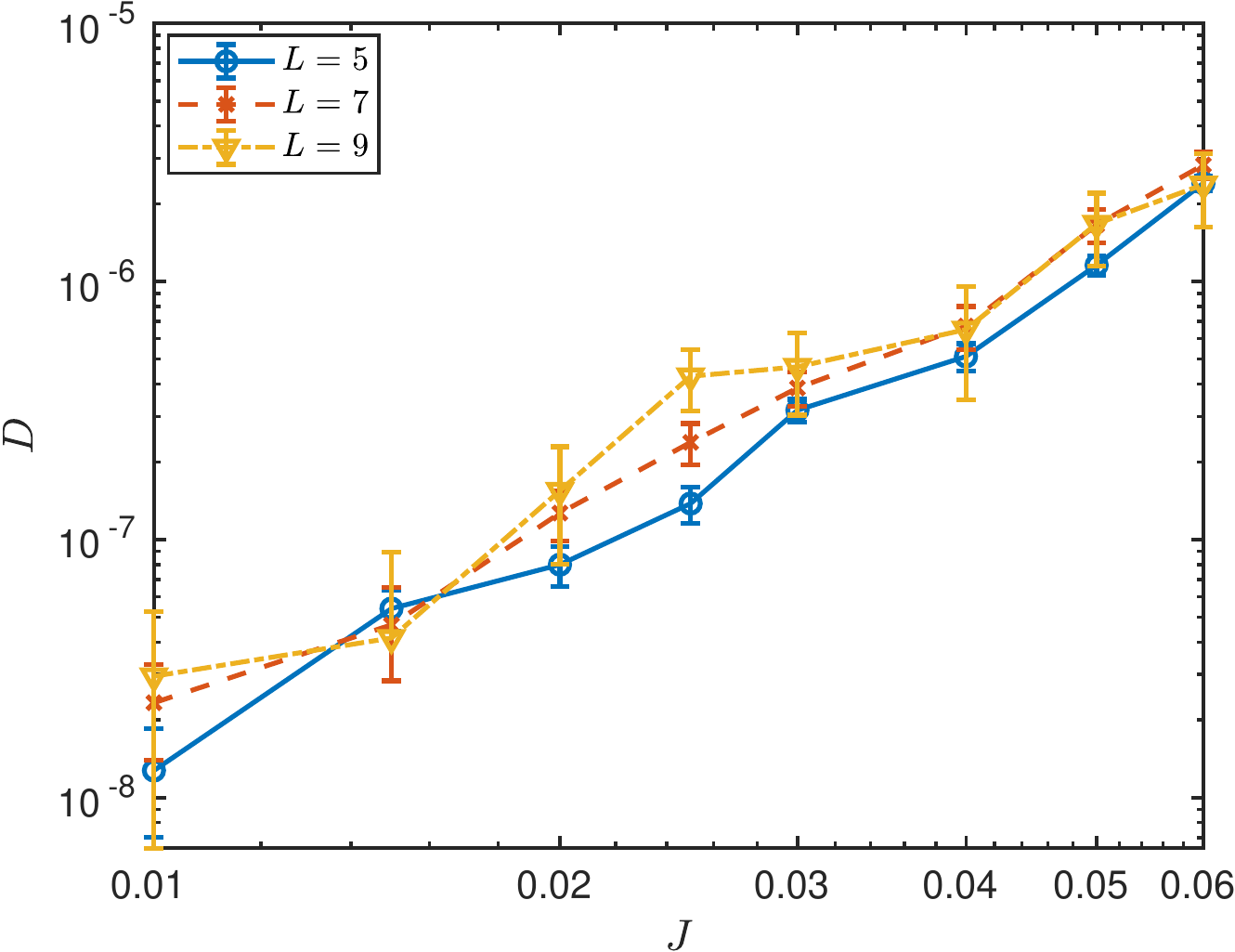}
 \caption{The diffusion coefficient $D$, while being almost independent from $L$, scales roughly as a power law with $J$.}
\label{fig:TWA_diffusion}
\end{figure}

To be sure that the dynamics is diffusive also at very small $J$, we have fitted $V(t)$ linearly for various $J$ and $L$ in the interval $10^4<t/\tau<3\cdot10^4$ (technical details on how to estimate the error are reported in Appendix~\ref{TWA:app}. We have thus estimated $D$ for various $J$ and $L$ (Fig.~\ref{fig:TWA_diffusion}). For $L\geq5$, $D$ exhibits only a weak dependence on $L$. Instead, for what concerns the $J$ dependence, $D$ roughly scales as a power law and most importantly always remains positive, signalling a thermal behaviour of the system. 

In conclusion the many-body dynamical localization we find is a genuine quantum effect without any classical counterpart. Classically the system always heats up to infinite temperature (as appropriate for a non-linear non-conservative many-body Hamiltonian system~\cite{Lichtenberg}).


\section{Conclusions and perspectives} \label{conc:sec}
We have found clear evidences that a kicked Bose-Hubbard model can show dynamical localization in the limit of large size (many-body dynamical localization). In this MBDL phase the local observables of the system do not thermalize to $T=\infty$, also in the limit of large system size. We have seen this fact studying the evolution of the static interacting part of the Hamiltonian $\hat{V}$. The MBDL phase involves all the spectrum and all the Floquet states violate ETH. We have shown this property by studying the distribution of the Floquet expectations of $\hat{V}$. In the MBDL phase the distribution remained {broad}, marking the absence of ETH and the related local equivalence of the Floquet states (typicality). 

In the MBDL phase the overlap with the initial state is very close to one and shows finite-size revivals at a time linearly scaling with the system size. On the opposite, in the ETH phase it decays to values exponentially small in the system size and does not show revivals. In the former case, the overlap behaviour corresponds to localization in the Hilbert space, while in the second to complete delocalization.

Looking at the entanglement entropy distribution of the Floquet states, we have seen that the average obeys a volume law both in the MBDL case and in the ETH case. On the opposite, the fluctuations over the distribution are linear in the system size in the former regular case, while they decay to zero exponentially fast in the latter thermalizing one (consistently with the ETH typicality). In the MBDL phase this gives rise to a time behaviour of the entanglement entropy of the following form. It increases linearly in time saturating to a value showing a volume-law scaling, much smaller than the thermal one. This linear increase in time is very similar to integrable systems but very different from MBL, where the increase in time is logarithmic.

Further insight can be gained by looking at the localization properties of the Floquet states. {In the MBDL phase these states are delocalized inside the $\hat{V}$ quasi-degenerate multiplet subspaces hybridized by the driving.} So, there is delocalization in sectors of the Hilbert space. This allows to interpret from one side the overlap near to one (the dynamics is localized to a subregion of the Hilbert space), and from the other side the linear increase in time of the entanglement entropy. This behaviour is indeed consistent with the ballistic propagation of excitations delocalized in space, which corresponds to delocalization inside the multiplet subspaces. {Being the dynamics localized across the quasi-degenerate subspaces of $\hat{V}$ allows also to understand why the increase of $V(t)$ is hindered.}

We have compared with the classical case (studied through TWA) and we have seen that classically the system always thermalizes to $T=\infty$. Ergodicity breaking is indeed a purely quantum phenomenon, as appropriate for dynamical localization.

We think it would be important to reach a better understanding of the MBDL phase and of the fact that there the dynamics is very similar to integrable cases, in terms of local observables and behaviour of the entanglement entropy. Probably, it is possible to construct an extensive set of integrals of motion (as done in~\cite{Anushya}) and this is a possible theme of future research. Another direction is the study of how long-range hopping can alter the regularity/ergodicity properties of the dynamics. From an experimental point of view, the phenomena we theoretically discuss in this work can be observed through the methods of the Harvard group~\cite{Lukin256,julian}.
%
\acknowledgements{We acknowledge fruitful discussions with B.~Altshuler, P.~Calabrese, M.~Heyl, F.~Iemini, G.~B. Mbeng, S.~Pappalardi, S.~Notarnicola and D.~Rossini. We thank Scuola Normale Superiore and D.~Rossini for the access to the GOLDRAKE cluster where part of the numerical calculations of this project were performed.}
\appendix
\section{Hilbert-space symmetrization} \label{symm:app}
In order to reduce the dimension of the Hilbert space, we exploit two symmetries of the model. We recall that the system is indeed invariant under two classes of transformations:
\begin{itemize}
 \item translations of $1$ site, {\it i.~e.} a relabelling of the sites $j\mapsto j+1\;(\text{mod}\, L)$;
 \item parity, {\it i.~e.} a relabelling $j\mapsto L-j+1$.
\end{itemize}
and, of course, also their compositions.

These symmetries are represented in the Hilbert space of the system through two operators $\hat{T}$ and $\hat{P}$ respectively. Since $\left[\hat T,\hat P\right]=0$, we can work in a basis of eigenstates of both operators. Their eigenvalues will be respectively
\begin{itemize}
 \item $T=\exp{i2\pi k/L}$, where $k$, the total momentum, is an integer in the interval $[-(L-1)/2,(L-1)/2]$ for odd $L$;
 \item $P=\pm1$.
\end{itemize}
The restriction on the possible eigenvalue is due to the constraint $\hat{T}^L=\hat{P}^2=\mathds{1}$

Since the transformations under study are symmetries of the Hamiltonian at every time, it will also be that $[\hat{U}_F,\hat T]=[\hat{U}_F,\hat P]=0$. Thus every Floquet states can be chosen to be also an eigenstate of $\hat T$ and $\hat P$. Since we are mainly interested in the dynamics starting from the uniform state, we will concern ourselves with the $k=0$ and $P=+1$ sector, which is the Hilbert space symmetrized w.r.t. translations and parity. In this appendix we aim to explain the obvious observations and lengthy case subdivision which we implemented in the code to construct first of all the symmetrized 
Hilbert space and then to calculate the hopping matrix elements in this basis.

First of all, we can notice that given any occupation string $\nb$, we can associate to it an element in the symmetrized Hilbert space
\begin{equation}
\label{eq:Fock-symm-formula}
 \ket{\tilde{\nb}}=\frac{1}{\sqrt{p(\nb)t(\nb)}}\sum_{a=0}^{p(\nb)-1}\sum_{b=0}^{t(\nb)-1} \hat{T}^a \hat{P}^b \ket{\nb}
\end{equation}
where $p(\nb)$ and $t(\nb)$ denotes the periodicity of the string $\nb$ w.r.t translations and parity respectively. In the following derivations it will be sometime useful to re-express
\begin{equation}
 \ket{\tilde{\nb}}=\sqrt{\frac{p(\nb)t(\nb)}{2L}}\sum_{a=0}^{L-1}\sum_{b=0}^{1} \hat{T}^a \hat{P}^b \ket{\nb}
\end{equation}
Using this observation it is very simple to implement an algorithm which build the symmetrized Hilbert space, while saving in the memory, for each state $\ket{\tilde{\nb}}$
\begin{itemize}
 \item one occupation string $\nb$ which can generate the state through Eq.~\eqref{eq:Fock-symm-formula},
 \item the translation period $t(\nb)$,
 \item the parity period $p(\nb)$.
\end{itemize}
we will refer to the dimension of the symmetrized space as $\dim\mathcal{H}_S\simeq \dim\mathcal{H}/2L$.

In order to calculate the matrix elements of $\hat{V}$, we can easily notice that it will be diagonal also in the $\ket{\tilde{\nb}}$ basis and $\hat{V}\ket{\tilde{\nb}}=V(\nb)\ket{\tilde{\nb}}$, with $\nb$ being any of the occupation string which generates $\ket{\tilde{\nb}}$.

For what concerns the matrix elements of $\hat{K}$ in this basis, it is a bit more cumbersome to work them out. One approach could be to store in memory also the unsymmetrized Hilbert space and the operator $K$ represented in the unsymmetrized basis. Defining then the matrix $\Pi$ of dimension $\dim\mathcal{H}_S \times \dim\mathcal{H}_S$, with elements
\begin{align}
 &\Pi_{\tilde{\nb},\nb'}=\braket{\nb|\tilde{\nb}}= \nonumber\\
 &\left\lbrace\begin{array}{ll}
   \frac{1}{\sqrt{p(\nb)t(\nb)}} & \text{if }\nb\text{ is a particular configuration of }\ket{\tilde{\nb}} \\
  0 & \text{otherwise} \\
\end{array}\right.
\end{align}
the matrix represented in the symmetrized basis is then given by
\begin{equation}
 \tilde{K}=\Pi K \Pi^\dag\,.
\end{equation}
%
\section{Two-states hybridization} \label{hybrid:app}
\begin{figure}
\begin{center}
	\begin{tabular}{c}
		\includegraphics[width=7cm]{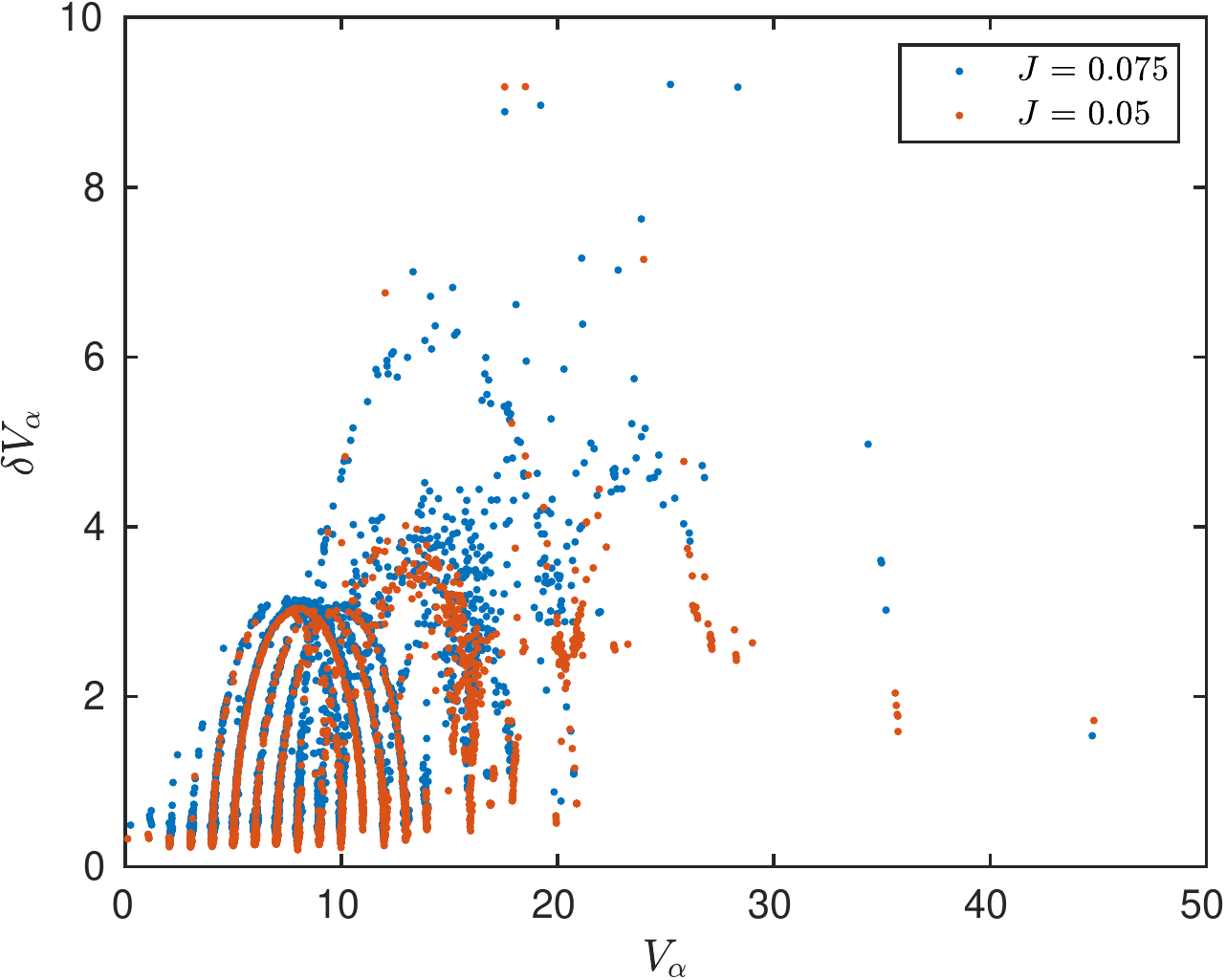}
	\end{tabular}
\end{center}
\caption{$\delta V_\alpha$ versus $V_\alpha$ for all the Floquet states $\ket{\phi_\alpha}$ for some value of $J$ that is intermediate between the MBDL and thermalizing phase. Numerical parameters: $L=10$.}
\label{VdV_intermediate:fig}
\end{figure}

In this section we study what happens at values of $J$ intermediate between the deep MBDL phase and the ergodic one. {In particular, we choose the value of $J=0.05$ which in Fig.~\ref{pdf:fig} appears to be the larger one consistent with dynamical localization.} The joint distribution of $V_\alpha$ and $\delta V_\alpha$ is reported in Fig.~\ref{VdV_intermediate:fig}.
A very simple two-states-hybridization model which allows to explain the fact that Floquet states are distributed along arcs. Indeed, the Floquet states $\ket{\phi_\alpha}$ along the arcs can be interpreted as an hybridization between two states $\ket{\psi_{1,2}}$ which are eigenstates of $\hat{V}$, $\hat{V}\ket{\psi_{1,2}}=V_{1,2}\ket{\psi_{1,2}}$. In fact, if $\ket{\phi_\alpha}=\gamma\ket{\psi_1}+\beta\ket{\psi_2}$, it is easy to show that
\begin{equation}
\label{eq:hybridizarc}
 \langle\delta V\rangle_{\phi_\alpha}= V_1 +(V_2-V_1) \sqrt{|\gamma|^2\left(1-|\gamma|^2\right)}\,.
\end{equation}

\begin{figure}
 \centering
 \includegraphics[width=70mm]{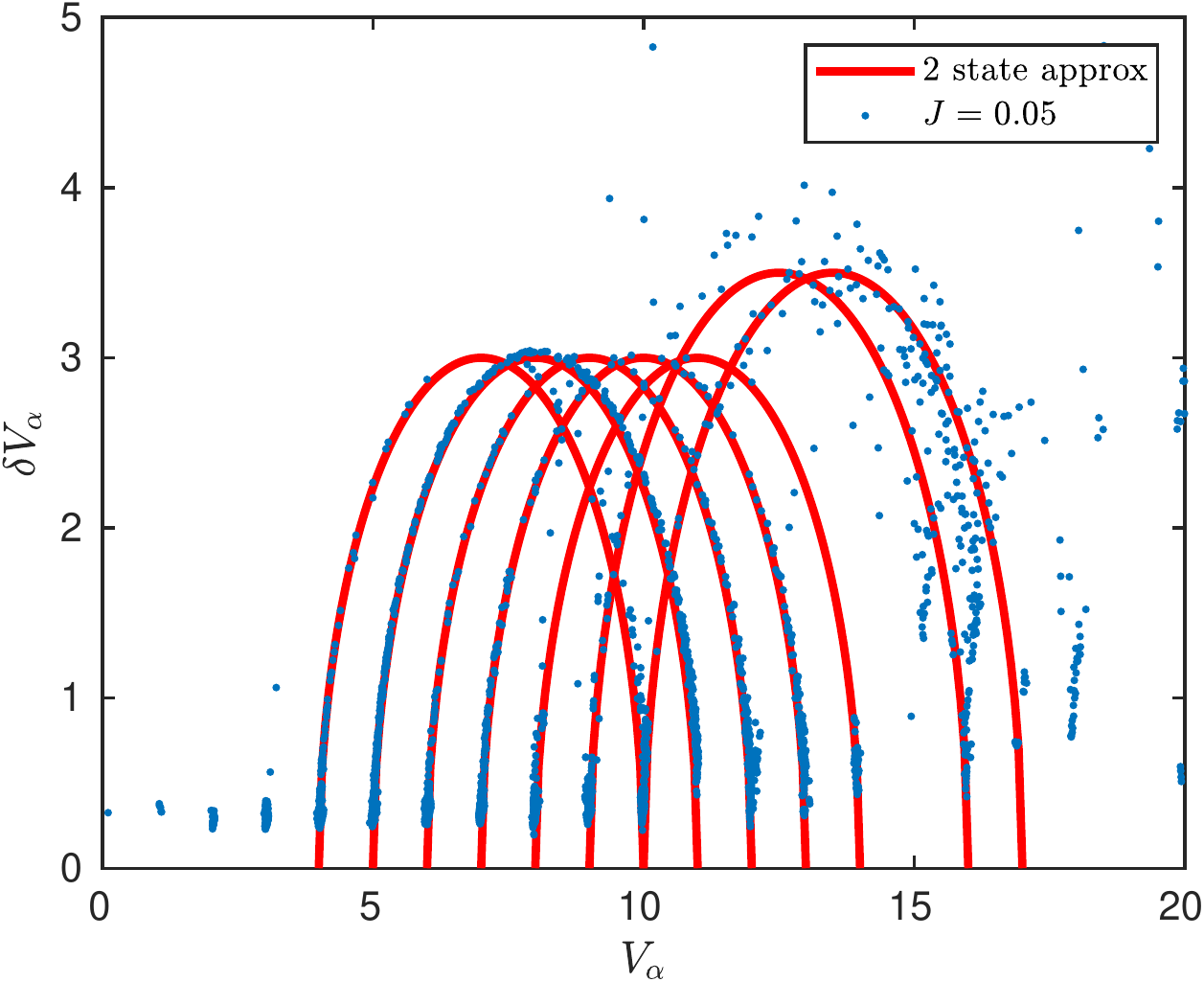}
 \caption{Analog of Fig.~\ref{VdV_joint_distribution:fig} for $J=0.05$. The arcs can be described as a hybridization of two eigenstates of $\hat V$ with different eigenvalue (2 states approximation). In fact, the arcs match very well the prediction in Eq.~\eqref{eq:hybridizarc}.}
\label{arcs:fig}
\end{figure}
This simple model describes the arcs quite well, as we show in Fig.~\ref{arcs:fig}.
This sequence of arcs might be microscopically explained by the activation of inelastic scattering processes where two sites with occupations $n=2$ and $n=3$ end up having occupation numbers $n=5$ and $n=0$.

%

\section{The truncated Wigner approximation for the kicked Bose-Hubbard model} \label{TWA:app}
In this appendix we describe the application of the TWA~\cite{POLKOVNIKOV_TWA} to our kicked Bose-Hubbard model.
A convenient semi-classical phase space for the bosonic modes on each site is given through coherent states
\begin{equation}
 \ket{\boldsymbol{\psi}}=\exp\left(\boldsymbol{\psi} \cdot \hat{\boldsymbol{a}}^\dag\right) \ket{\boldsymbol{0}}
\end{equation}
where $\hat{\boldsymbol{a}}^\dag=(\ha_1,\ha_2,\cdots,\ha_L)$ and $\boldsymbol{\psi}\in\mathbb{C}^L$. It is thus possible to define the Wigner quasi-distribution in $\boldsymbol{\psi}$ space associated to a density matrix $\hat{\rho}$ as
\begin{align}
 &W(\bp)=\nonumber\\
 &\frac{1}{2^L}\int d\be d\be^* \Braket{\bp-\frac{\be}{2}|\hat{\rho}|\bp+\frac{\be}{2}} 
 e^{ -|\bp|^2 -\frac{|\be|^2}{4}}e^{\frac{1}{2}(\be^* \bp -\be \bp^*)}\,.
\end{align}
through the Wigner quasi-distribution it is indeed possible to compute the expectation value of an operator $\hat{O}$ on a given state, simply by
\begin{equation}
 \text{Tr}\left(\hat{\rho} \hat{O}\right)=\int d\bp d\bp^* W(\bp) O_W(\bp)
\end{equation}
where $O_W$ is the Weyl symbol of $\hat{O}$ defined as%
\begin{align}
 &O_W(\bp)=\nonumber\\
 &\frac{1}{2^L}\int d\be d\be^* \Braket{\bp-\frac{\be}{2}|\hat{O}|\bp+\frac{\be}{2}} 
 e^{ -|\bp|^2 -\frac{|\be|^2}{4}}e^{\frac{1}{2}(\be^* \bp -\be \bp^*)}\,.
\end{align}
The way of computing the averages justifies the interpretation of $W$ as a distribution function in real space~
\footnote{However one should be careful in this analogy since, unlike a classical probability distribution function, the Wigner distribution is not always positive, {\it i.~e.} it may have regions in $\bp$ space where $W(\bp)<0$. This is why it is often referred to as {\it quasi-}distribution}.
The TWA amounts to initialize the system in the quantum Wigner distribution corresponding to the given initial state and make evolve the variables $\boldsymbol{\psi}$ in the phase space with a classical effective Hamiltonian which is the lowest order in an expansion in $\hbar$~\cite{POLKOVNIKOV_TWA}. More precisely, the expectations of the observables in this approximation scheme are computed as
\begin{equation}
\label{eq:TWA_average}
 \langle \hat{O}(t) \rangle = \int d\bp_0 W(\bp_0) O_W\left(\bp_{cl}(t)\rvert_{\bp(0)=\bp_0}\right)
\end{equation}
where $\bp(t)$ satisfies the equation (denoting with $H_W$ the Weyl symbol of the Hamiltonian)
\begin{equation} \label{eqmot:eqn}
 i\frac{d}{dt}\psi_j = \frac{\partial }{\partial \psi_j^*}H_W(t)\,.
\end{equation}
For reader's convenience we report in Table~\ref{tab:weyl-symbols} the Weyl symbols useful to construct $H_W(t)$ in our case.

%
\begin{table}
\caption{Weyl symbols of the most relevant operators for the kicked Bose-Hubbard model. We add (see last line) that the Weyl symbol the product of two operators acting on different sites is the product of the Weyl operators.}
\[
 \begin{array}{ll}
  \toprule
  \hat{O} & O_W(\bp)\\
  \midrule
  \ha_j		& \psi_j	\\
  \ha_j^\dag	& \psi_j^*	\\
  \hn_j		& |\psi_j|^2-1/2\\
  \hn_j(\hn_j-1)& |\psi_j|^4 -2 |\psi_j|^2+1/2\\
  \midrule
\hat{A}_j\hat{B}_{j'}, \; j\neq j' & A_W(\bp) B_W(\bp)\\
  \bottomrule
 \end{array}
\]
\label{tab:weyl-symbols}
\end{table}
%
Since the on-site repulsion term and the hopping term are never present at the same time in the Hamiltonian, it is straightforward to integrate Eq.~\eqref{eqmot:eqn}. Between two subsequent kicks we have (up to a global phase) 
\begin{equation}
 \psi_j\left((m+1)\tau^-\right)=\psi_j\left(m\tau^+\right)\exp\left(iU\tau \psi_j \left|\psi_j(m\tau^+)\right|^2\right)
\end{equation}
while the evolution given by the kick can be expressed in terms of a propagator $G(j-j')$ which can be easily obtained exploiting translation invariance and applying the Fourier transform to the variables $\psi_j$
\begin{align}
 &\psi_j\left(m\tau^+\right)=\sum_{j'}G(j-j') \psi_{j'}\left(m\tau^-\right)\quad{\rm with}\nonumber\\
 &G(j-j')= \frac{1}{L} \sum_k e^{2\pi i \frac{k(j-j')}{L}}\exp\left(-2iJ \cos\frac{2\pi k}{L}\right)\,.
\end{align}
Quite nicely, in the limit $L\to\infty$ we can express the propagator as
\begin{equation}
 G(j-j')=(-i)^{|j-j'|}\mathcal{J}_{|j-j'|}(2J)\,,
\end{equation}
where $\mathcal{J}_n (2J)$ are the cylindrical Bessel functions of order $n\geq 0$.

For what concerns the Wigner distribution of the uniform state, it is simply given by the product
$W(\bp)= \prod_j W_\nu(\psi_j)$ where $W_\nu(\psi_j)$ is the Wigner function of the $\nu$-th number eigenstate of a bosonic mode and can be expressed in terms of Laguerre polynomial $L_n$ as (see for example~\cite{gerry_knight_2004})
\begin{equation}
\label{eq:wigner-distr}
 W_\nu(\psi)=\frac{2}{\pi}(-1)^\nu L_\nu(4|\psi|^2)\exp(-2|\psi|^2)\,.
\end{equation}

The numerical implementation of the method depicted above is quite straightforward via Montecarlo, apart from one point. If the Wigner function were a positive function, we could generate a random $\bp$ with a probability distribution given by $W(\bp)$, then evolve $\bp$ in time, while keeping trace of the observables of interest $O_W(\bp(t))$. The average over the various $\bp$ would then correspond exactly to the expression~\ref{eq:TWA_average}.
The subtle question is how to deal with the non-positive definite Wigner function in Eq.~\eqref{eq:wigner-distr}. Fixing the number of samples, the approach that minimizes the error is the following. We can rewrite Eq.~\eqref{eq:TWA_average} as follows
\begin{align}
 &\langle \hat{O}(t) \rangle =\nonumber\\
 &Z_\nu \int d\bp_0 \frac{\left|W(\bp_0)\right|}{Z_\nu} \text{sign}\left(W(\bp_0)\right)O_W\left(\bp_{cl}(t)\rvert_{\bp(0)=\bp_0}\right)
\end{align}
where $Z_\nu=\left[\int d^2 \psi |W_\nu(\psi)|\right]^L$ is just a normalization factor. We will thus employ $\left|W(\bp_0)\right|/Z_\nu$ as a probability measure, while sampling the factor on the right. Finally we can evaluate $Z_\nu$ by straightforward numerical integration.

In order to generate a random $\bp$ according to the above probability, we employ a Markov-chain Montecarlo~\cite{frenkel2001understanding} to determine each $|\psi_j|$. In order to do so, we start the Markov chain with $|\bp|=0$. Then at each step of the chain we select randomly a $j$, $1\le j\le L$ and a $\delta \psi$ uniformly distributed in the interval $[-\Delta, \Delta]$ with $\Delta=\mathcal{O}(1)$. So we propose the substitution $|\psi_j|\to |\psi_j|+\delta \psi$. If $|\psi_j|+\delta \psi< 0$ the proposal is immediately rejected, otherwise the acceptance probability is calculated
\begin{equation}
 P_{\rm acc}=\left|\frac{W_\nu(|\psi_j|+\delta \psi)(|\psi_j|+\delta\psi)}{W_\nu(|\psi_j|)|\psi_j|}\right|\,.
\end{equation}
The move is then accepted with probability $P_{\rm acc}$, where it is intended that if $P_{\rm acc}>1$ the move is always accepted. Applying the detailed balance principle, it is straightforward to show that the described Markov chain generates the desired distribution $\prod_j|\psi_j|W(|\psi_j|)$.
Finally, before calculating the time evolution starting from $\bp$, we multiply each $|\psi_j|$ by a random phase.

When we evaluate the diffusion coefficient $D$ (see Fig.~\ref{fig:TWA_diffusion}) we must correctly estimate the error. In order to do that, we must take into account the correlation between the energy at time $t$ with the energies at nearby times. For this reason we have employed a jackknife resampling technique~\cite{efron1982jackknife}. So, we remove one Montecarlo (say the $j$-th one) sample and we fit the means of the remaining data linearly, obtaining in this way an estimate $D_j$ for the diffusion constant. This procedure is repeated varying $j$ over all the $\mathcal{N}$ samples. The jackknife estimate on the error of $D$ is then given by
\begin{equation}
  \delta D=\sqrt{\mathcal{N}-1}
  \sqrt{\frac{1}{\mathcal{N}}\sum_j D_j^2 -\left(\frac{1}{\mathcal{N}}\sum_j D_j\right)^2}\,.
\end{equation}
This procedure is important since the error evaluated directly from the fitting procedure would be some order of magnitudes smaller.
\section{An excited initial state} \label{02:app}
\begin{figure}
 \begin{center}
   \includegraphics[width=7cm]{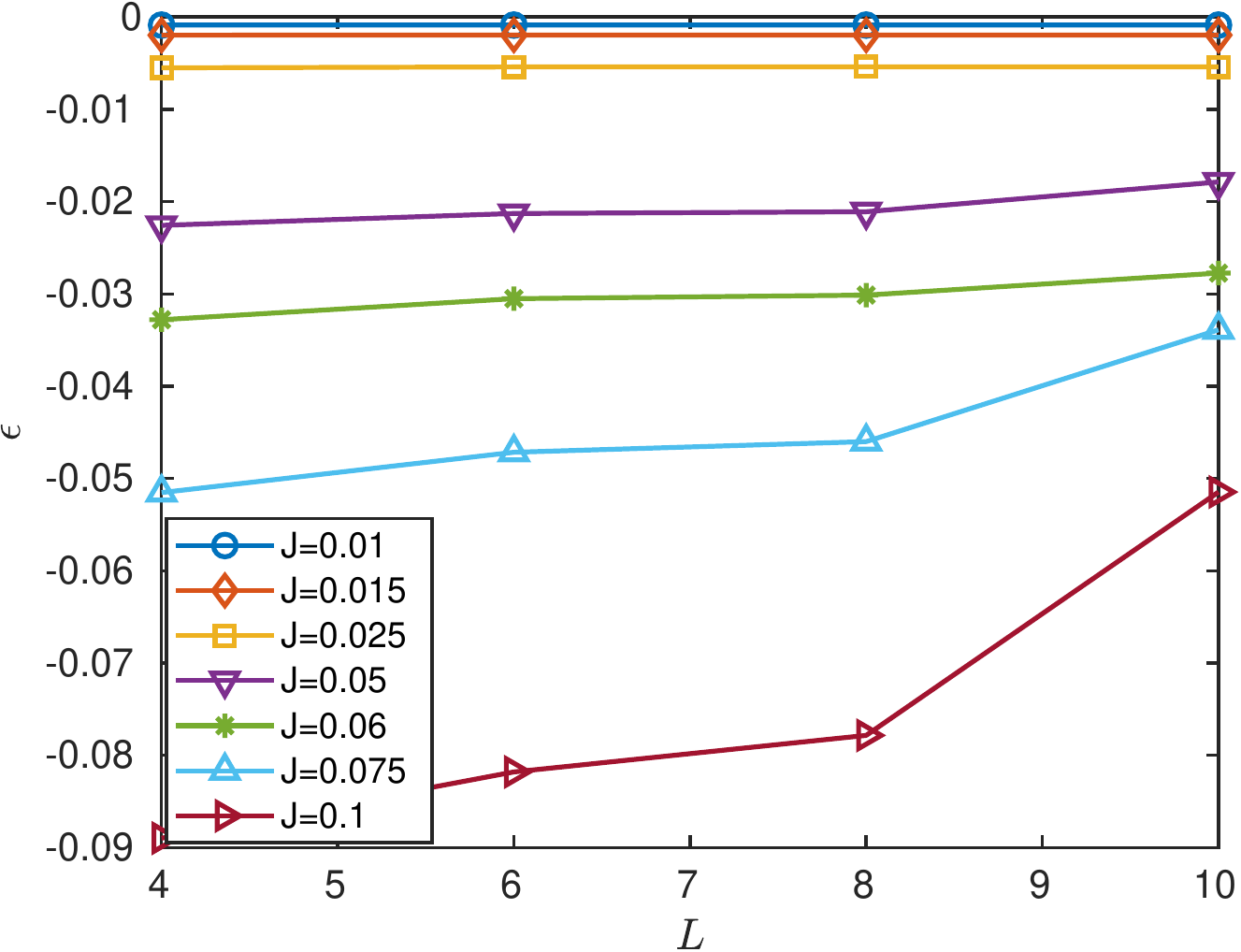}\\
   \includegraphics[width=7cm]{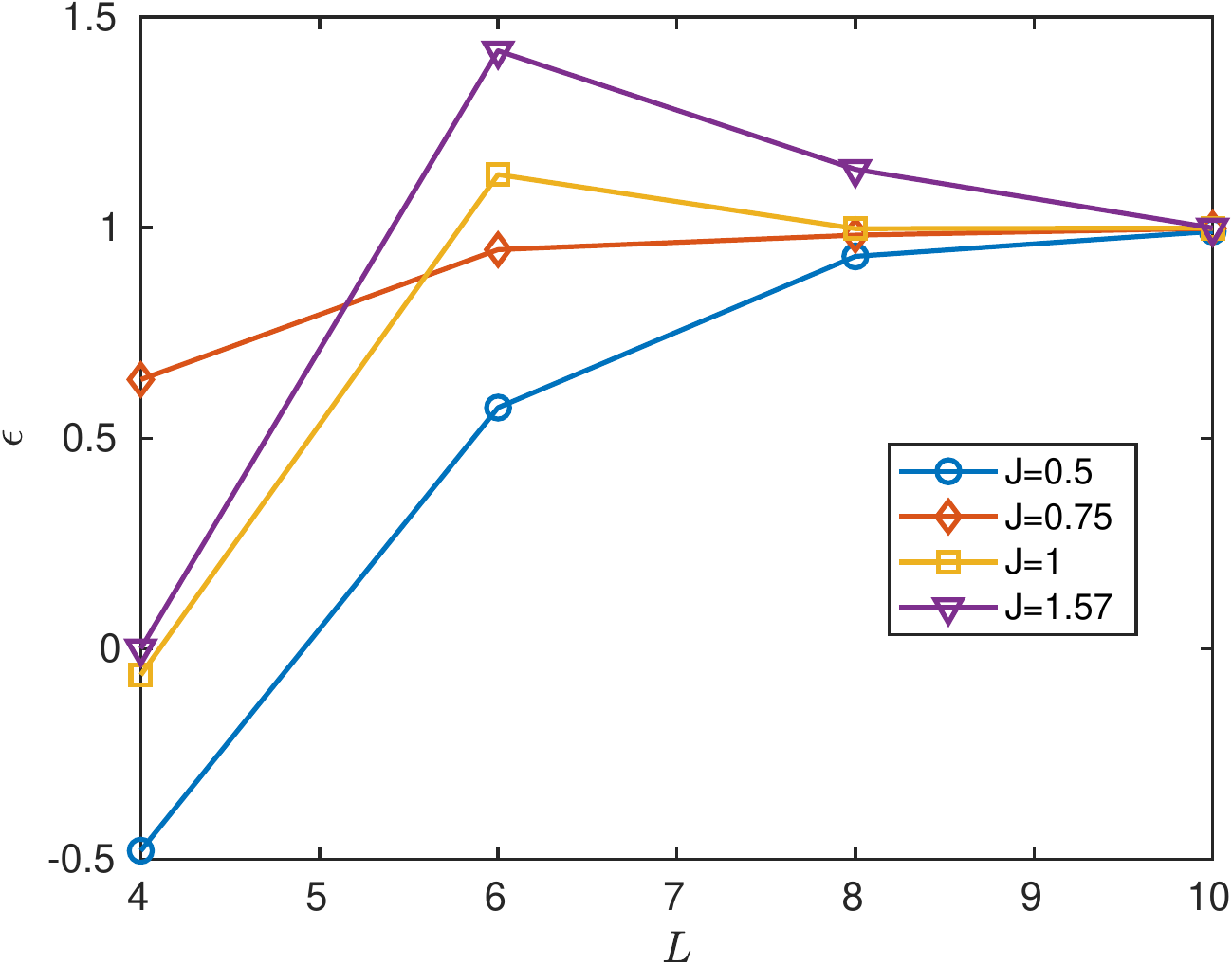}
 \end{center}
 \caption{$\epsilon$ versus $L$ for the initial state $\ket{\psi_{02}(0)}$.}
 \label{eps_vs_L2:fig}
\end{figure}

\begin{figure}
 \begin{center}
   \begin{tabular}{c}
    \includegraphics[width=7cm]{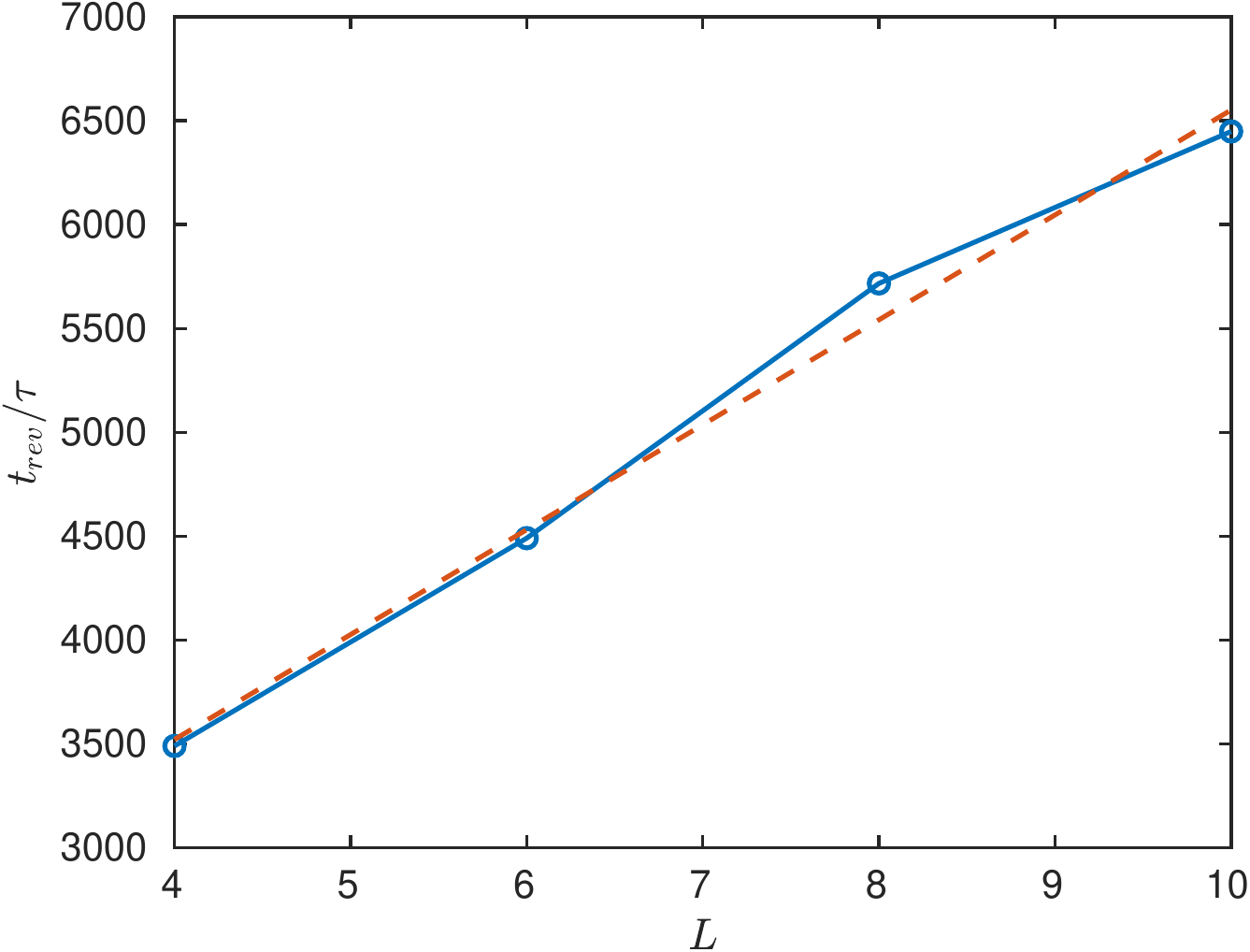}
   \end{tabular}
 \end{center}
 \caption{Size-scaling of the revival time for a MBDL case ($J=0.01$) and excited initial state with initial state $\ket{\psi_{02}(0)}$. Notice the approximately linear dependence on $L$.}
 \label{scalingtime2:fig}
\end{figure}

\begin{figure}
 \begin{center}
   \begin{tabular}{c}
    \includegraphics[width=7cm]{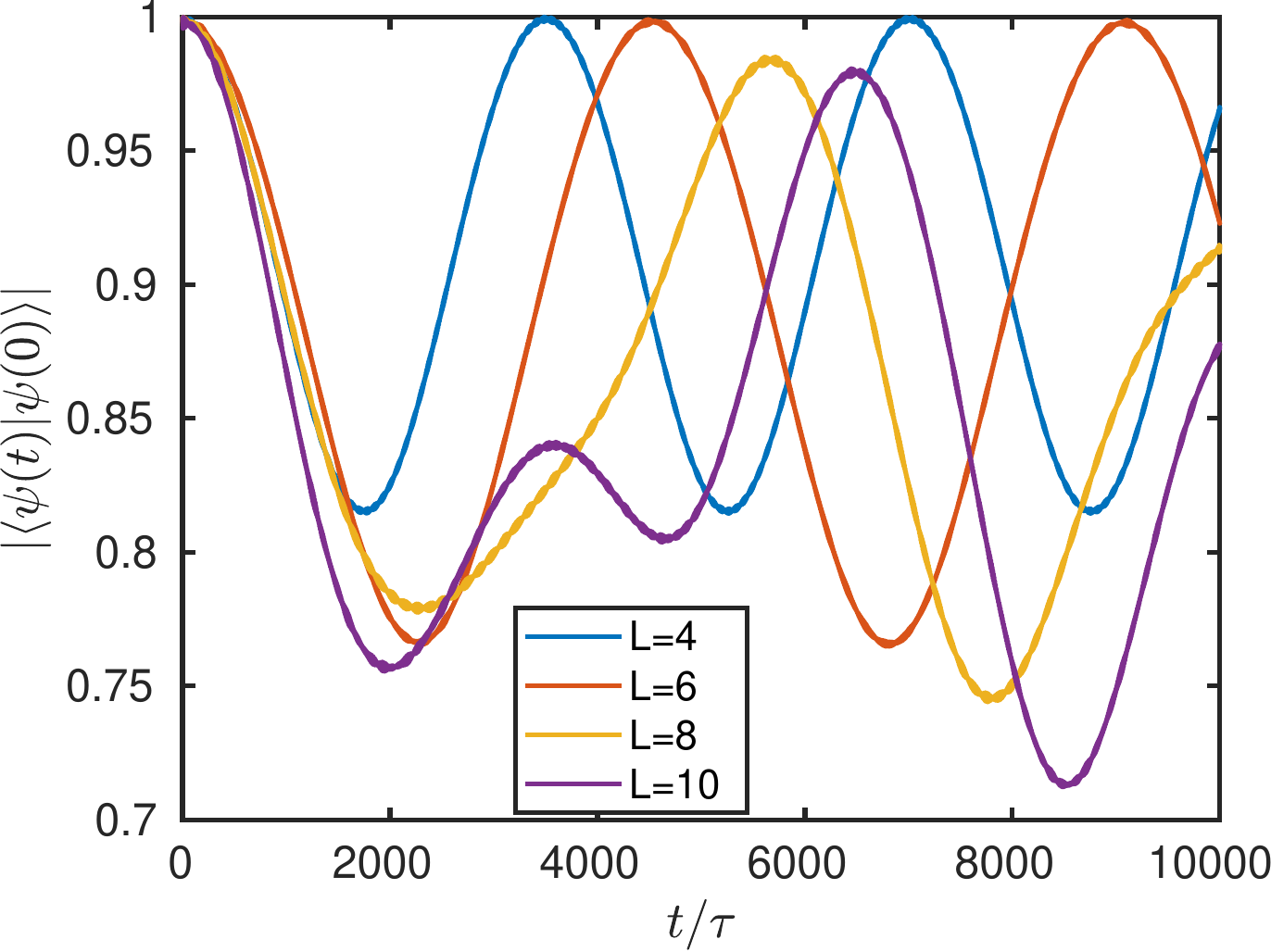}
   \end{tabular}
 \end{center}
 \caption{Overlap dynamics in a dynamically localized case ($J=0.01$) and excited initial state.}
 \label{nonlinari2:fig}
\end{figure}

Here we consider as initial state 
{
\begin{widetext}
\begin{equation}
  \ket{\psi_{02}(0)}=\frac{1}{\sqrt{2}}\left(\;\ket{2}\otimes\ket{0}\otimes\ket{2}\otimes\ldots\otimes\ket{0}\;+\;\ket{0}\otimes\ket{2}\otimes\ket{0}\otimes\ldots\otimes\ket{2}\;\right)\,.
\end{equation}
\end{widetext}
We take $L$ even so that this is a state with filling factor $\nu=1$.} This state has a finite initial excitation-energy density with respect to the ground initial state with all $n_j=1$. Nevertheless, if we initialize the dynamics with this excited state, we see the same evidences of a transition from thermalization to MBDL as we saw for the uniform state. In particular, Fig.~\ref{eps_vs_L2:fig} is the equivalent of Fig.~\ref{eps_vs_L:fig}, Fig.~\ref{scalingtime2:fig} the equivalent of Fig.~\ref{revival_scaling:fig}, and Fig.~\ref{nonlinari2:fig} the equivalent of the upper panel of Fig.~\ref{nonlinari:fig}.
%

\begin{thebibliography}{10}
\bibitem{Nota1}
\newblock {For $V_{T=\infty}$ we either use its exact finite-size value, computed directly as the trace of $\hat{V}$ in the symmetric subspace, or for $L>10$ we approximate $V_{T=\infty}/L\simeq\nu^2$ which becomes exact in the $L\to\infty$ limit, as can be seen by exchanging the fixed-number ensemble with a grand canonical one (fixed chemical-potential).}
\end{thebibliography}

%
\end{document}